\documentclass[twocolumn,epjc3]{svjour3}
\usepackage{graphicx}
\graphicspath{{./picture/}}
\usepackage{empheq}
\usepackage{amssymb}
\usepackage{mathtools}
\usepackage{commath}
\usepackage{verbatim}
\usepackage{xcolor}
\usepackage{soul}
\usepackage{booktabs}
\usepackage{amsmath}
\usepackage{hhline}
\usepackage{url}
\usepackage[numbers, sort&compress]{natbib}
\usepackage{epsfig}
\usepackage{lscape}
\usepackage{mathrsfs}
\usepackage{nicefrac} 
\usepackage{upgreek}
\usepackage{bbm}
\usepackage{psfrag}
\usepackage{fontawesome5}
\usepackage{float}

\usepackage{diagbox}
\usepackage{makecell}
\usepackage[compat=1.1.0]{tikz-feynman}
\usetikzlibrary{calc}

\RequirePackage[T1]{fontenc}
\smartqed  

\RequirePackage{graphicx}
\RequirePackage{mathptmx}      
\RequirePackage{flushend}
\RequirePackage[numbers,sort&compress]{natbib}
\RequirePackage[colorlinks,citecolor=blue,urlcolor=blue,linkcolor=blue]{hyperref}

\journalname{Eur. Phys. J. C}

\DeclareMathAlphabet{\mathpzc}{OT1}{pzc}{m}{it}

\newcommand{\cdec}{c}

\usepackage{hyperref}
\newcommand{\A}{b}
\newcommand{\Aalpha}{\A_{\alpha}}

\newcommand{\B}{\hat{\Gamma}}

\DeclareMathAlphabet{\mathcal}{OMS}{cmsy}{m}{n}
\newcommand{\XX}{\mathcal{A}}
\newcommand{\YY}{\mathcal{B}}
\newcommand{\N}{X}
\newcommand{\n}{N_{\rm min}}

\newcommand{\pN}{{\rm p}}
\newcommand{\GammaN}{\Gamma}
\newcommand{\lambdaN}{\lambda}
\newcommand{\M}{m}
\newcommand{\GammaNinv}{\Gamma_{\rm inv}}
\newcommand{\GammaNvis}{\B_{\rm vis}}
\newcommand{\Bnbeta}{\B_\beta}

\newcommand{\MaD}[1]{\textcolor{black}{ #1}}
\newcommand{\new}[1]{\textcolor{black}{ #1}}

\begin{document}
\vspace{-1.5cm}

\title{{\bf \boldmath New Particles at the Z-Pole:}}
\subtitle{\Huge Tera-Z factories as discovery and precision machines}

\vspace{-1.5cm}

\title{New Particles at the Z-Pole: \subtitle{Tera-Z factories as discovery and precision machines}}

\author{Marco Drewes\thanksref{e1,addr1,addr2}
        \and
        Juraj Klari\'{c}\thanksref{e2,addr3}
        \and
        Yuan-Zhen Li\thanksref{e3,addr1}
}
\thankstext{e1}{e-mail: marco.drewes@uclouvain.be}
\thankstext{e2}{e-mail: jklaric@phy.hr}
\thankstext{e3}{e-mail: yuanzhen.li@uclouvain.be}

\institute{
UCLouvain, Centre for Cosmology, Particle Physics and Phenomenology, CP3, Chemin du Cyclotron 2, 1348 Louvain-la-Neuve, Belgium
\label{addr1}
           \and
           Physik–Department, Technische Universität München, D-85748 Garching, Germany \label{addr2}
           \and
           Department of Physics, Faculty of Science, University of Zagreb, 10000~Zagreb,~Croatia \label{addr3}
}

\date{Received: date / Accepted: date}

\maketitle
\begin{abstract}
  \noindent 
Several proposed future lepton colliders are capable of producing trillions of Z-bosons, including FCC-ee, CEPC, LEP3 and LEP-Z. Such Tera-Z factories can discover new elementary particles with couplings to the Z-boson that are orders of magnitude smaller than current bounds. For couplings near the currently excluded parameter regions they could produce sufficiently large samples to study the new particles' properties in detail, thereby serving as both a discovery and precision machine in one. Using simple analytic estimates, we quantify the dependence of the expected event yield in long-lived particle searches on the number of produced Z-bosons and on the detector dimensions. From this, we derive estimates for both the discovery reach and the measurement precision attainable at such facilities. While the precision of such estimates of course falls short of proper simulations, the analytic approach is suitable for a quick assessment of the sensitivity for a given design. We illustrate this with two examples, heavy neutral leptons and axion-like particles. Under optimistic assumptions, these could be produced in the millions and billions, respectively, effectively turning future lepton colliders into exotics factories.
\end{abstract}

\section{Introduction}

\begin{table*}
	\centering
	\begin{tabular}{|c||c|c|c|c|}
    \hline
		collider & FCC-ee & CEPC & LEP3 & LEP-Z  \\
		\hline
		total $N_Z$ & $6\times 10^{12}$ \ \cite{FCC:2025lpp} & $4.1\times 10^{12}$ \ \cite{Ai:2025cpj} & $1.7\times 10^{12}$ \ \cite{Anastopoulos:2025jyh} & $6\times 10^{12}$ \cite{Drewes:2025ohy} \\
        \hline
	\end{tabular} 
    \caption{Expected total number of Z-bosons for different colliders under current plans.}
	\label{tab:NZ}
\end{table*}

Several proposed concepts for future lepton colliders foresee an extended operation at a collision energy of 91 GeV to produce trillions of Z-bosons as so-called Tera-Z factories \cite{Altmann:2025feg,Arduini:2947728,deBlas:2025gyz}, including the FCC-ee \cite{FCC:2018evy,FCC:2025lpp}, CEPC \cite{CEPCStudyGroup:2018ghi,CEPCStudyGroup:2023quu,CEPCStudyGroup:2025kmw,Ai:2025cpj}, LEP3 \cite{Anastopoulos:2025jyh} and LEP-Z \cite{Drewes:2025ohy}. 
The precise number of produced Z-bosons $N_Z$ depends not only on the machine's design, but also on the schedule of operation, and in particular on how much integrated luminosity is dedicated to the Z-pole run. Table \ref{tab:NZ} summarises the numbers foreseen with the current planning.\footnote{It should be remarked that FCC-ee and CEPC are very similar machines, so the difference in their $N_Z$ reported in the table could be alleviated by assigning more or less integrated luminosity to the Z-pole run. Similarly, the number of Z-bosons that can be produced per annum at LEP3 and LEP-Z is similar; the difference in $N_Z$ comes from the fact that LEP-Z can run for longer at the Z-pole.} 
To what degree the shares of integrated luminosity at different collision energies can be adapted 
after the machine has been built 
is thus partly a political decision,
and partly depends on many technical details that go beyond the scope of this work.\footnote{For instance, operating at the Higgs boson or top quark threshold requires a different RF system than the Z-pole, hence the schedule cannot be simply changed on the fly. Moreover, a crucial factor that has to be decided in advance is the number of interaction points.} 
We therefore refrain from making sensitivity predictions for any particular collider in the following, but instead study what can be done for a given $N_Z$, treating the numbers in table \ref{tab:NZ} as indicative for specific machines.

Such large samples of Z-bosons can advance our understanding of fundamental physics in different ways. 
Focussing on signatures of physics beyond the Standard Model (SM) of particle physics, one can broadly distinguish two avenues.  
\begin{itemize}
    \item \emph{Indirect signatures.}
    If the mass $\M$ of a new elementary particle exceeds the  collision energy, it can only leave an indirect trace in observables as a virtual mediator. In this case deviations from the SM can be described in the framework of effective field theories (EFTs) involving SM fields.
    The huge number of events that can be achieved with $N_Z\sim 10^{12}$ permits searches for such deviations through very precise measurements of known processes as well as searches for very rare events. 
    The uncertainties in most electroweak and flavour observables  are dominated by backgrounds and statistics, see \cite{wilkinson_2025_nf05j-xsq05,FCC:2025lpp,Ai:2024nmn,blondel_2025_zv2qx-xk656}, and the errors scale as $\propto 1/\sqrt{N_Z}$.
Moreover, experimental systematic uncertainties are often statistics-limited because one can use the huge event samples to quantitatively measure the cause of the potential systematic biases and uncertainties.
\item \emph{Discovery of new elementary particles}.
It is possible that new elementary particles with masses $\M$ below the LHC collision energy have escaped discovery because they rarely interact with ordinary matter.
 This can either occur in minimal models featuring isolated SM gauge singlets or in more complex theories in which a hidden or dark sector couples to the SM only through so-called portals \cite{Alekhin:2015byh,Curtin:2018mvb,Beacham:2019nyx,Agrawal:2021dbo}. We quantify the suppression of the pertinent interactions relative to the SM gauge couplings by a set of small parameters $\{ \upepsilon_i\}$.  
 The number of new particles produced in Z-boson decays is directly proportional to $N_Z$. 
 To assess the discovery potential of a given facility, it is crucial to understand for which range of values of the  $\{ \upepsilon_i\}$ a statistically significant number of events can be observed in a given detector. When studying the properties of newly discovered particles, it is also important to estimate how many events may be seen within this sensitivity region.
\end{itemize}

We focus on the second avenue, i.e., the discovery of new elementary particles. 
If these decay into SM particles with a sizeable branching ratio, then they often have a comparably long proper lifetime, owing to the smallness of the $\{ \upepsilon_i\}$.  
This has triggered considerable effort into searches for long-lived particles (LLPs) at the LHC \cite{Alimena:2019zri} and other facilities \cite{Beacham:2019nyx,Agrawal:2021dbo,Antel:2023hkf}. 
It is well-known that lepton colliders have an immense discovery potential for LLPs, and numerous studies have been performed \cite{Blondel:2022qqo,FCC:2025lpp,Altmann:2025feg,Ai:2025cpj}.

The purpose of the present work is not to propose any new search or to improve the accuracy of the forecasts made in earlier studies. 
Instead, we use very simple analytic formulae to estimate and illustrate the discovery potential and capability for precision measurements that can be achieved for given values of $N_Z$ and the detector dimensions. 
In section \ref{sec:GenericLLPanalytic} we phrase these in a model-independent way,
in section \ref{sec:examples} we apply them to specific examples, namely { heavy neutral leptons} (HNLs) and axion-like particles (ALPs).

While the accuracy of our analytic estimates is clearly below that of extensive simulations that have been performed in the literature, they represent a simple and quick framework to assess the sensitivity region for discoveries as well as the precision at which particle properties can be measured as functions of $N_Z$ and the detector dimensions. This can help to facilitate quick estimates of these quantities in discussions about the design and schedule of future colliders and detectors. 
We provide a code that quickly generates the sensitivity curves displayed in this work and can be extended to other models at \href{https://github.com/liyuanzhen98/LLPatTeraZ}{\faGithub\ LLPatTeraZ}.

\section{Analytic formulae for generic LLPs}\label{sec:GenericLLPanalytic}

Let us consider a new particle $\N$ with a mass $\M$ below the electroweak scale that couples to Z-bosons with an effective interaction strength that is considerably feebler than the SM weak interaction. We quantify the resulting suppression of the $\N$-production cross section in Z-decays by a small parameter $\upepsilon_{\rm pro}\ll 1$. 
We further introduce a second small parameter $\upepsilon_{\rm dec}$ that characterises the small probability of $\N$ decaying into observable SM final states. 
The $\upepsilon_{\rm pro}$ and $\upepsilon_{\rm dec}$ can  be thought of as an agnostic effective parametrisation of Wilson coefficients that govern the strength of $\N$-interactions in an EFT framework. Fundamentally they are typically proportional to the square of some coupling constant or mixing angle. 
In general the coefficients $\upepsilon_{\rm pro}$ and $ \upepsilon_{\rm dec}$ should be thought of as independent because production and decay may be mediated by different interactions. 
We will consider separately the important special case where the same coupling governs production and decay of $\N$, i.e., $\upepsilon_{\rm pro} =  \upepsilon_{\rm dec} = \upepsilon$.

The discovery potential of a given Z-factory can be characterised by the minimal $\upepsilon_{\rm pro}$ and $\upepsilon_{\rm dec}$ that lead to a desired minimal number of observable number of events $\n$. Our goal is to explore the dependence of this sensitivity region on $N_Z$ and the detector geometry. 
Fundamentally the sensitivity is limited by three factors.
    \begin{itemize}
    \item[1)]\label{Nprodlimit}  
    \emph{Integrated luminosity:}
  If the number of produced $\N$-particles 
  $N_{\rm prod} \propto \upepsilon_{\rm prod} N_Z$ is smaller than one, then a discovery is impossible. 
\item[2)] \emph{Detector size:}
If the $\N$ decay length in the laboratory $\lambdaN$ exceeds some critical scale $l_1$,
the fraction of observable decays within the fiducial detector volume becomes so small that the number of observed decays $N_{\rm obs}$ drops below one. 
\new{For far detectors, there is an additional requirement that $\lambdaN$ exceeds a minimal value $l_0$ related to the distance of the detector from the collision point.}
\item[3)] \emph{Backgrounds:} With $\N$-decays generally being rare events, a discovery hinges on a good control over the backgrounds. 
\MaD{In LLP searches many backgrounds (in particular those from SM particles from the primary vertex) can be suppressed by appropriate cuts if $\lambdaN$ exceeds some critical value $l_0$ that depends on the details of the detector and the search.}
\end{itemize}

 The number of $\N$ that is produced along with a given state $\XX_\alpha$ in decays of on-shell Z-bosons is
          \begin{eqnarray}
        N_{\rm prod}^{\XX_\alpha}
        =
        N_Z \   \upepsilon_{\rm pro} \Aalpha    
    \end{eqnarray} 
where $\Aalpha$  parameterises 
the branching as ${\rm Br}(Z\to \XX_\alpha \ \N) = \upepsilon_{\rm pro}\Aalpha$. The total number of produced $\N$ is 
        \begin{eqnarray}\label{Nprodgeneral}
        N_{\rm prod} = \sum_\alpha 
        N_{\rm prod}^{\XX_\alpha}
           \equiv 
           N_Z \   \upepsilon_{\rm pro} \A.
    \end{eqnarray} 
    
The typical decay length of $\N$ in the laboratory frame $\lambdaN$ is related to its decay rate $\GammaN$  via the Lorentz boost factor $\pN / \M$,
\begin{equation}\label{DecayLength}
\lambdaN = \frac{\pN / \M}{\GammaN},
 \end{equation}
 with $\pN$ the magnitude of the $\N$'s three-momentum and $\M$ its mass. 
In order for the $\N$-momentum $\pN$ in \eqref{DecayLength} to be monochromatic, the state $\XX_\alpha$ observed in the detector (which may consist of several particles after hadronisation) must originate from a single particle at the vertex, i.e., $\N$ is initially produced in a $1\to2$ decay. This is the case in the two examples discussed in Section \ref{sec:examples}. 

In general $\N$ can decay into a set of observable final states $\YY_\beta$ 
as well as invisible final states.
The total rate is
    \begin{eqnarray} \label{GammaLLP}
\GammaN = 
\GammaNinv +  \upepsilon_{\rm dec} \sum_\beta \Bnbeta
\equiv
\GammaNinv +  \upepsilon_{\rm dec}  \GammaNvis, 
\end{eqnarray}
where $\GammaNinv$ is the decay rate into invisible final states and the partial decay widths $\upepsilon_{\rm dec} \Bnbeta$
parameterise the
branching ratio ${\rm Br}(\N \to \YY_\beta) = \upepsilon_{\rm dec} \Bnbeta/\GammaN$. 
In the following we adapt the nomenclature \emph{invisible final states}  with respect to a specific search. That is, they can consist of particles that are truly invisible for a given detector (e.g.~neutrinos or sterile dark sector particles), but the term may also refer to states that are just not included in a given search. 

\begin{figure*}[ht]
    \centering
    \begin{tikzpicture}
  \begin{feynman}
    \vertex (a);                  
    \vertex [right=2.5cm of a, dot, label=95:$\upepsilon_{\text{pro}}$] (b) {};
    
    \vertex [above left= 1.8 cm of a] (i1) {$e^-$};
    \vertex [below left=1.8 cm of a] (i2) {$e^+$};
    
    \vertex [above right=2.1cm of b] (f1) {$\XX_\alpha$};
    \vertex [below right=2.1cm of b] (f2) {$\N$};

    \diagram* {
      (i1) -- [fermion] (a),
      (i2) -- [anti fermion] (a),
      (a) -- [boson, edge label'=$Z$] (b),
      (b) -- [plain] (f1),
      (b) -- [plain] (f2),
    };

    \vertex [right=6cm of a] (x_in) {$\N$}; 
    \vertex [right=3cm of x_in, dot, label=95:$\upepsilon_{\text{dec}}$] (v_dec) {};
    
    \diagram* {
      (x_in) -- [plain] (v_dec),
    };
    \def\coneLen{2.8}    
    \def\coneRad{0.8}    
    \def\conePersp{0.25} 
    \filldraw[fill=gray!20, draw=black] 
        (v_dec.center) -- 
        ++(\coneLen, \coneRad) -- 
        ++(0, -2*\coneRad) -- 
        cycle;
    \filldraw[fill=gray!40, draw=black] 
        ($(v_dec) + (\coneLen, 0)$) ellipse (\conePersp cm and \coneRad cm);
    \node [right] at ($(v_dec) + (\coneLen + \conePersp, 0)$) {$\YY_\beta$};
  \end{feynman}
\end{tikzpicture}
    \caption{\emph{Left}: Production of an LLP $\N$ along with  $\XX_\alpha$.
    \emph{Right}: Subsequent decay of $\N$ into a multiparticle state $\YY_\beta$.
    }
    \label{fig:feynman-1}
\end{figure*}
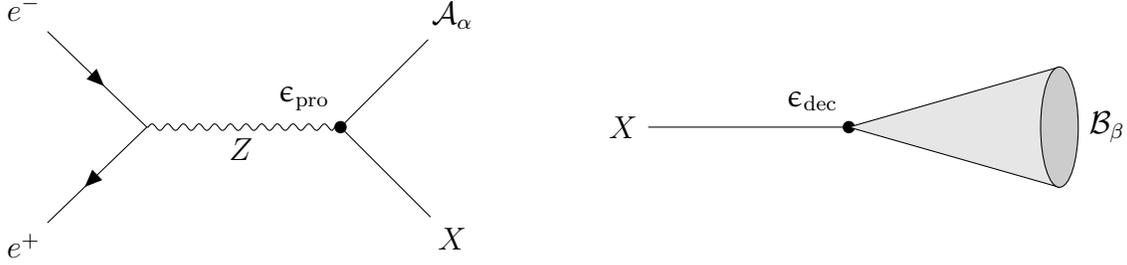

The number of observable events 
with $l_0 < \lambdaN < l_1$ 
in which $\N$ was produced along with $\XX_\alpha$ and decays into $\YY_\beta$ is then\footnote{
In our approach we factorise the entire process $e^+ e^- \to \XX_\alpha \YY_\beta$ into $e^+ e^- \to \XX_\alpha \N $ and $\N \to \YY_\beta$ by putting the $\N$-particle on-shell, thus neglecting quantum interferences and off-shell contributions to the $\N$-propagation. Examples where this may not be justified include U(1) extensions of the SM in which the Z' is mass-degenerate with the Z-boson or HNLs with degenerate masses, see \cite{LoChiatto:2024guj,Blondel:2021mss,Antusch:2023jsa,Antusch:2023nqd} and references therein. However, if the time scale associated with coherent oscillations is either much faster or much slower than the lifetime, \eqref{NobsGeneral} can still be used when introducing effective parameters that model the impact of coherent effects, such as non-integer values of the parameter $\cdec$ in \eqref{GammaDefinition} \cite{Drewes:2022rsk,Drewes:2024pad}.}
    \begin{eqnarray}\label{NobsGeneral}
    N_{\rm obs}^{\XX_\alpha, \YY_\beta} 
    &=& N_{\rm prod}^{\XX_\alpha}
     \ \left[e^{-l_0/\lambdaN} - e^{-l_1/\lambdaN} \right] \ {\rm Br}(\N\to  \YY_\beta)
    \nonumber\\
    &=&    \upepsilon_{\rm pro} \ N_Z \ \Aalpha \ \left[e^{-l_0/\lambdaN} - e^{-l_1/\lambdaN} \right] 
    \frac{\upepsilon_{\rm dec}\Bnbeta}{
\GammaN}.
    \end{eqnarray}
The total number of observed events  $N_{\rm obs}$ is obtained by summing over $\alpha$ and $\beta$.
We  assume 100\% reconstruction efficiency for all final states. Relaxing this assumption could be achieved by introducing appropriate efficiency factors in \eqref{NobsGeneral}.

When studying the properties of the new particles $\N$, { important observables are the branching ratios into specific final states} $N_{\rm obs}^{\XX_\alpha, \YY_\beta}/N_{\rm obs}$. The statistical uncertainty on their measurements can be estimated as \cite{Antusch:2017pkq}
\begin{eqnarray}\label{precision}
    \frac{\delta(N_{\rm obs}^{\XX_\alpha, \YY_\beta}/N_{\rm obs})}{N_{\rm obs}^{\XX_\alpha, \YY_\beta}/N_{\rm obs}}
    =
    \sqrt{\frac{1}{N_{\rm obs}^{\XX_\alpha, \YY_\beta}}-\frac{1}{N_{\rm obs}}}.
\end{eqnarray}

In the following we identify sensitivity regions in terms the coefficients $\upepsilon_{\rm pro}$, $\upepsilon_{\rm dec}$, $\upepsilon$ based on the requirement to see at least $\n$ events. For a discovery $\n$ is usually of order one, a common choice is $\n > 4$. However, $\n$ can more generally be identified with the number of events needed to push the statistical uncertainty in a given parameter below a desired value \cite{Cowan:2010js}. 
A summary of the ranges of validity of the various analytic expressions is given in Table \ref{OverviewTable}.

\paragraph{Limitations due to the integrated luminosity.}
   The integrated luminosity at the Z-pole poses the hardest limit, as $N_{\rm prod} < 1 $ implies that not even the most futuristic detectors have any chance of discovery. Evidently $N_{\rm prod} \propto N_Z$ in the case under consideration here, c.f.~\eqref{Nprodgeneral}, highlighting the importance of Tera-Z factories.
Requesting $N_{
\rm prod} > \n $ implies the hard limit
\begin{eqnarray}\label{eq:general}
    \upepsilon_{\rm pro} \ > \ \frac{\n}{N_Z \A }
\end{eqnarray}
which holds in all cases. 
Assuming that all visible final states $\YY_\beta$ are detected, but
allowing for decays into invisible states 
implies a stronger limit,
\begin{eqnarray}\label{Labskaus}
    \upepsilon_{\rm pro} \ > \ \frac{\n}{N_Z \A } \Big/ \left(1 - \frac{\GammaNinv}{\GammaN}\right),
\end{eqnarray}
which applies even in the case of an infinitely large detector \MaD{($l_1/\lambdaN\to\infty$) and in the limit $l_0/\lambdaN\to0$}. 
When either the detector geometry or the 
\MaD{suppression of backgrounds} 
requires that we can only count events that occur in a limited interval $l_0 < \lambdaN < l_1$, this imposes a stronger lower bound
\begin{eqnarray}\label{eprodtot}
\upepsilon_{\rm pro} 
> \frac{\n}{N_Z \A}
        \left[
    \left(1 - \frac{\GammaNinv}{\GammaN}\right)
    \left(\exp(-l_0/\lambdaN) - \exp(-l_1/\lambdaN)\right) 
    \right]^{-1}.    
\end{eqnarray}

For the special case $\upepsilon_{\rm pro} =  \upepsilon_{\rm dec} = \upepsilon$
we find an analytic formula only in the limit $l_0/\lambdaN, \ \lambdaN/l_1 \to 0$, 
\begin{eqnarray}\label{sameWWNZ}
    \upepsilon \ > \  \frac{\n}{2N_Z \A}\left(
1 + \sqrt{1 + 4\frac{ \A N_Z }{\n}\frac{\GammaNinv}{\GammaNvis}}
    \right),
\end{eqnarray}
analogously to \eqref{Labskaus}.
As expected, when $\N$ primarily decays into observable final states, \eqref{sameWWNZ} scales as $\propto \n/N_Z$, but it scales as $\propto \sqrt{\n/N_Z}$ if $\N$ decays predominantly into the hidden sector.
In the limit of very small $\upepsilon$ (implying that $\N$ predominantly decays into invisible states) one finds for an infinitely large detector ($l_1/\lambdaN \to \infty$)
\begin{eqnarray}\label{Borka}
\upepsilon \ \ > \ \ e^{l_0
\frac{\GammaNinv}{\pN/\M}/2} \sqrt{\frac{\n }{N_Z \A}\frac{\GammaNinv}{\GammaNvis}} 
\
\underset{l_0/\lambdaN \to 0}{\simeq}  
\
\sqrt{\frac{\n }{N_Z \A}\frac{\GammaNinv}{\GammaNvis}}.
\end{eqnarray}
In the limit where invisible decays can be neglected 
($\GammaNinv/\GammaNvis \ll \upepsilon$) 
one finds (still 
$l_1/\lambdaN \to \infty$)
\begin{eqnarray}\label{Kenny}
    \upepsilon \ > \ -\frac{\pN /\M}{ \GammaNvis l_0} W_0\left(-\frac{ l_0  \GammaNvis }{\pN/\M} \frac{\n }{N_Z \A}\right)
    \
\underset{l_0/\lambdaN \to 0}{\simeq}  
\
    \frac{\n }{N_Z \A},
\end{eqnarray}
with $W_s$ the $s$-branch of the Lambert W-function.
The last step in \eqref{Borka} and \eqref{Kenny} applies for  $l_0/\lambdaN\to 0$ and can be obtained from the corresponding limiting cases of \eqref{sameWWNZ}.

\paragraph{Limitations due to the detector dimensions.}
To investigate the limitations due to the detector geometry, 
we consider cylindrical detectors with a fiducial volume of diameter $d_{\rm cyl}$ and length $l_{\rm cyl}$.
For the specific case of LLPs produced from Z-boson decays at rest with an approximately isotropic angular distribution, 
it turns out that setting 
      \begin{eqnarray}\label{l1def}
         l_1 = \frac{1}{2} (3/2)^{1/3} 
 d_{\rm cyl}^{2/3} l_{\rm cyl}^{1/3}
    \end{eqnarray}
in \eqref{NobsGeneral} captures the dependence of the sensitivity region on the detector geometry surprisingly well \cite{Drewes:2022rsk}.\footnote{For a spherical detector $l_1$ would simply be the radius of the sphere.}

For searches in the main detectors a detection does not require a minimal displacement\MaD{;} in this context $l_0$ can be used to estimate the impact of a cut on the displacement that removes SM backgrounds, see below.
For far detectors, however, the parameter $l_0$ should be related to their distance from the collision point into account.
Most far detectors do not have a full $4\pi$ solid angle coverage.\footnote{The dimensions of the largest realistic $4\pi$ detector are determined by the cavern size \cite{Chrzaszcz:2020emg}.} However, for the isotropic angular distribution considerer here, \eqref{NobsGeneral} and all subsequent equations can still be used when reducing the number of events by an appropriately chosen factor representing the effective fraction of the covered solid angle.

We first consider the case that $\upepsilon_{\rm pro}$ and $\upepsilon_{\rm dec}$ are independent. Then the effect of a finite detector size on the lower bound on $\upepsilon_{\rm pro}$ is already included in \eqref{eprodtot}. Finding a lower bound on $\upepsilon_{\rm dec}$ is in general not possible analytically. 
In the limit of very small $\upepsilon_{\rm dec}$ (implying that $\N$ predominantly decays into invisible states, 
$\GammaNinv/\GammaNvis \gg \upepsilon_{\rm dec}$) one finds
\begin{eqnarray}\label{IntoTheDark}
    \upepsilon_{\rm pro}\upepsilon_{\rm dec} \ &>& \ \frac{\GammaNinv}{\GammaNvis}\frac{\n }{N_Z \A} \nonumber \\
    &&\Big/
    \left(\exp\left(-l_0 
    \frac{\GammaNinv }{\pN/\M }
    \right) - \exp\left(-l_1 \frac{ \GammaNinv }{\pN/\M}\right)\right)
\end{eqnarray}
In the limit where all final states can be observed 
($\GammaNinv/\GammaNvis \ll \upepsilon_{\rm dec}$)
one finds for $l_0/\lambda\to 0$
\begin{eqnarray}\label{Heimdall}
    \upepsilon_{\rm dec} \ > \ 
    -   \frac{\pN/\M}{\GammaNvis l_1}
    \ln\left(
    1 - \frac{\n}{\A N_Z \upepsilon_{\rm pro} }
    \right) .
\end{eqnarray}
For very long-lived $\N$ ($\lambdaN \gg l_1 \gg l_0$) we find an analytic solution for arbitrary $\GammaNinv/\GammaNvis$,
\begin{eqnarray}\label{Methusalix}
\upepsilon_{\rm dec} \upepsilon_{\rm pro} \ > \ \frac{\n}{N_Z \A}   \frac{\pN/\M}{\GammaNvis l_1}.
\end{eqnarray}
Note that \eqref{Methusalix} can be obtained from \eqref{Heimdall} in the limit $\n \ll N_{\rm prod}$, but is more general than this limit: \eqref{Heimdall} assumes a negligible branching ratio into invisible final states and is valid for arbitrary $l_1/\lambdaN$;
\eqref{Methusalix} holds for any $\GammaNinv/\GammaN$, but requires very long-lived $\N$ ($\lambdaN \gg l_1$).

In the special case $\upepsilon_{\rm pro} =  \upepsilon_{\rm dec} = \upepsilon$ the bounds \eqref{IntoTheDark} 
and 
\eqref{Methusalix}
hold with the replacement $\upepsilon_{\rm pro}\upepsilon_{\rm dec} \to \upepsilon^2$ on the l.h.s..

The resulting limitations on proposed main detectors \cite{IDEAStudyGroup:2025gbt,CEPCStudyGroup:2025kmw} can partially be mitigated by extra instrumentation and far detectors \cite{Wang:2019xvx,Chrzaszcz:2020emg,Tian:2022rsi}.  
\begin{table*}[ht]
\centering
\renewcommand{\arraystretch}{1.5} 
\begin{small}
\begin{tabular}{|l||c|c|c|}
\hline
\diagbox{decay length}{decay modes} & 
\begin{tabular}{c}
primarily invisible\\
$\GammaNinv/\GammaNvis \gg \upepsilon_{\rm dec}$ 
\end{tabular}
& \begin{tabular}{c}
primarily visible\\
$\GammaNinv/\GammaNvis \ll \upepsilon_{\rm dec}$ 
\end{tabular} & general \\
\hline
\hline
$l_0, \lambdaN \ll l_1$
& 
\eqref{eprodtot}, \eqref{Borka}, \eqref{IntoTheDark}
& 
\eqref{eprodtot}, 
\eqref{Kenny},  
\eqref{Heimdall2},  \eqref{Kenny2}
& \eqref{eprodtot}\\
\hline
$l_0 \ll \lambdaN \ll l_1$
&  
\begin{tabular}{c} 
\eqref{Labskaus}, \eqref{eprodtot}, \eqref{sameWWNZ}, \\ \eqref{Borka}, \eqref{IntoTheDark}
\end{tabular}
& 
\begin{tabular}{c} \eqref{eq:general}, \eqref{Labskaus}, \eqref{eprodtot},\eqref{sameWWNZ}, \\
\eqref{Kenny},  \eqref{Heimdall}, 
\eqref{Heimdall2},  \eqref{Kenny2}
\end{tabular}
& 
\eqref{Labskaus}, \eqref{eprodtot}, \eqref{sameWWNZ}
\\
\hline
$l_0 \ll \lambdaN, l_1$
& \eqref{eprodtot}, \eqref{IntoTheDark} & \eqref{eprodtot}, \eqref{Heimdall} & \eqref{eprodtot} \\
\hline
$l_0 \ll l_1 \ll \lambda$
& \eqref{eprodtot}, \eqref{IntoTheDark}, \eqref{Methusalix}  & \eqref{eprodtot}, \eqref{Heimdall}, \eqref{Methusalix} & 
\eqref{eprodtot}, \eqref{Methusalix}
\\
\hline
general & \eqref{eprodtot}, \eqref{IntoTheDark}  & \eqref{eprodtot} &  \eqref{eprodtot}  \\
\hline
\end{tabular}
\end{small}
\caption{\label{OverviewTable}Validity of the analytic formulae for LLP sensitivity limitations in various limiting cases.
Here $\lambdaN$ refers to the decay length in the lab frame \eqref{DecayLength}.
$l_1$ characterises the detector dimensions, for a cylindrical detector it is given by \eqref{l1def}.   
For searches with the main detectors the parameter $l_0$  can be used to estimate the impact of a cut on the displacement, for far detectors it should be identified with their distance from the collision point. 
Whenever $\upepsilon$ appears without label, the expression assumes $\upepsilon_{\rm pro} = \upepsilon_{\rm dec} = \upepsilon$.
In this case  \eqref{IntoTheDark} 
and 
\eqref{Methusalix}
hold with the replacement $\upepsilon_{\rm pro}\upepsilon_{\rm dec} \to \upepsilon^2$.}
\end{table*}

\paragraph{Limitations due to backgrounds or far detectors.}
The parameter $l_0$ can either mimic the effect of a cut needed to remove backgrounds or indicate the distance of a far detector from the collision point.  
The lower bound on $\upepsilon_{\rm pro}$ due to $l_0$ is again already given by \eqref{eprodtot} without further assumptions.
The bound on $\upepsilon_{\rm pro}$ when $\N$ predominantly decays invisibly ($\GammaNinv/\GammaNvis \gg \upepsilon_{\rm dec}$) is given by \eqref{IntoTheDark}.
Physically the presence of the exponential involving $l_0$ signifies that only a small fraction of the observable decays occur in the region where backgrounds can be neglected (or inside a far detector) even if we consider an infinitely large detector ($l_1/\lambdaN\to\infty$). 
For $\GammaNinv/\GammaNvis \ll \upepsilon_{\rm dec}$
the requirement that enough $\N$ decay outside the region haunted by backgrounds imposes an upper bound
\begin{eqnarray}\label{Heimdall2}
    \upepsilon_{\rm dec} <  
    \frac{\pN/\M}{\GammaNvis l_0}
    \ln\left(
    \frac{\A N_Z \upepsilon_{\rm pro} }{\n}
    \right) .
\end{eqnarray}
The reason why we find a lower bound \eqref{IntoTheDark} when $\N$ primarily decays invisibly and an upper bound \eqref{Heimdall2} when it predominantly decays visibly can be understood intuitively.
In the latter case the decay length \eqref{DecayLength} is inversely proportional to  $\upepsilon_{\rm dec}$, increasing this parameter implies that more and more decays occur in the background-haunted region of radius $l_0$ around the collision point. Hence, requiring a given number of events $\n$ outside this region imposes an upper bound on $\upepsilon_{\rm dec}$.
In the former case the impact of $\upepsilon_{\rm dec}$ on the decay length is negligible, and increasing this parameter only enhances the fraction of visible decays (compared to invisible ones). Therefore $ N_{\rm obs} \sim \upepsilon_{\rm dec}$, and requiring $N_{\rm obs} > \n$ imposes a lower bound on $\upepsilon_{\rm dec}$.

Also for the special case $\upepsilon_{\rm pro} =  \upepsilon_{\rm dec} = \upepsilon$ analytic expressions can be found in these two limiting cases. 
For $\N$ that predominantly decays into invisible final states ($\GammaNinv/\GammaNvis \gg \upepsilon_{\rm dec}$) there is a lower bound
on $\upepsilon$ given by \eqref{Borka}.
For negligible branching ratio into invisible final states ($\GammaNinv/\GammaNvis \ll \upepsilon_{\rm dec}$)
there is an upper bound on $\upepsilon$ to assure that a sufficient number of $\N$ decay outside the region plagued with backgrounds \MaD{in the main detectors or inside a far detector placed at a distance $l_0$},
\begin{eqnarray}\label{Kenny2}
    \upepsilon \ &&< \ -\frac{\pN}{\M \GammaNvis l_0} W_{-1}\left(-\frac{\GammaNvis l_0}{\pN/\M } \frac{\n }{N_Z \A}\right)
   \nonumber \\
    &&\underset{\frac{l_0}{\lambdaN}\frac{\n}{N_{\rm obs}}\upepsilon\ll 1}{\simeq} 
    \
    -\frac{\pN}{\M \GammaNvis l_0} \ln\left(\frac{ \GammaNvis l_0}{\pN/\M} \frac{\n }{N_Z \A}\right),
\end{eqnarray}
which envelopes the sensitivity region of an infinitely large detector together with \eqref{Kenny}.
$W_s$ is again the Lambert function.

Note that the requirement to 
\MaD{suppress}
backgrounds does not forbid a discovery, it just means that our simplified formulae overestimate the sensitivity. For instance, the precision estimate \eqref{precision} is only valid if the uncertainty is entirely statistics-limited, which is usually not the case in the presence of backgrounds. 
Since backgrounds can be understood and removed by improved algorithms, triggers etc., the choice of $l_0$ \MaD{for searches in the main detectors} is neither limited by fundamental physics nor engineering.
\new{In this sense one could say that, in principle, restrictions due to $l_1$ -- which is directly related to the detector geometry -- are harder than those due to $l_0$, as sizeable changes in the former require modifying the detector size, 
while the latter may be improved by employing a better performing detector technology.}\footnote{
\MaD{In practice, the fact that the dependence of $N_{\rm obs}$ on $l_0$ tends to be stronger than that on $l_1$ (exponential vs power law) limits the improvement that can be achieved, so that the best strategy is usually to combine displaced and prompt searches to cover the entire parameter region.}}
\MaD{In contrast, for far detectors, where $l_0$ signifies their distance from the collision point, a sizeable reduction of $l_0$ does require extra instrumentation or even civil engineering.}

\section{Illustrative examples}\label{sec:examples}

In the following we illustrate the sensitivity of Tera-Z factories in two well-motivated benchmark models involving HNLs and ALPs.

\subsection{Heavy neutral leptons}
Extensions of the SM by heavy right-handed Majorana neutrinos solve several open problems in particle physics and cosmology \cite{Drewes:2013gca}.
Most notably, they can explain the light neutrino properties \cite{Esteban:2024eli} through the seesaw mechanism \cite{Minkowski:1977sc,Glashow:1979nm,Gell-Mann:1979vob,Mohapatra:1979ia,Yanagida:1980xy,Schechter:1980gr},
generate the matter-antimatter asymmetry of the universe \cite{Canetti:2012zc} via  leptogenesis \cite{Fukugita:1986hr} and are viable Dark Matter candidates \cite{Dodelson:1993je} (see e.g.~\cite{Bodeker:2020ghk} and \cite{Boyarsky:2018tvu} for reviews). 
Well-motivated and technically natural models with heavy neutrino masses below the TeV scale exist \cite{Agrawal:2021dbo}, 
and can be probed in collider experiments \cite{Atre:2009rg,Deppisch:2015qwa,Antusch:2016ejd,Abdullahi:2022jlv}.
 For instance, the Neutrino Minimal Standard Model ($\nu$MSM) \cite{Asaka:2005an,Asaka:2005pn} represents a benchmark scenario in which all these problems can be solved simultaneously \cite{Canetti:2012kh,Ghiglieri:2020ulj}, and that has served as a well-motivated benchmark for searches at Tera-Z factories \cite{Blondel:2014bra,Blondel:2022qqo,FCC:2025lpp,Altmann:2025feg} (see also \cite{Antusch:2015mia,Antusch:2016vyf,Barducci:2020icf,Shen:2022ffi,Chakraborty:2018khw}).
 On a more general basis, heavy neutrinos can simultaneously explain the neutrino masses and the matter-antimatter asymmetry of the universe in the entire parameter region accessible to displaced vertex searches at Tera-Z factories \cite{Drewes:2021nqr}. 

From the viewpoint of collider phenomenology, the heavy right-handed neutrinos appear as a type of HNL of mass $M$ with a coupling to the weak force that is suppressed by the elements of a matrix $\theta$ which characterise their mixing with the different SM neutrinos $\nu_{L \alpha}$, with $\alpha = e,\mu,\tau$.
Realistic models that can explain the light neutrino oscillation data require at least two flavours of HNLs. However, many aspects of the collider phenomenology can effectively be parameterised by the Lagrangian \cite{Atre:2009rg}
\begin{eqnarray}
 \mathcal L
&\supset&
- \frac{m_W}{v} \overline N \theta^*_\alpha \gamma^\mu e_{L \alpha} W^+_\mu
- \frac{m_Z}{\sqrt 2 v} \overline N \theta^*_\alpha \gamma^\mu \nu_{L \alpha} Z_\mu \nonumber \\
&& - \frac{M}{v} \theta_\alpha h \overline{\nu_L}_\alpha N
+ \text{h.c.},
\label{PhenoModelLagrandian}
\end{eqnarray}
with $m_Z$, $m_W$ the weak gauge boson masses and $v\simeq 174$ GeV the Higgs field vacuum expectation value.

\begin{figure*}[ht]
    \centering
    \begin{tikzpicture}
  \begin{feynman}
    \vertex (a);                  
    \vertex [right=2.5cm of a, dot, label=95:$U^2$] (b) {};
    
    \vertex [above left= 1.8 cm of a] (i1) {$e^-$};
    \vertex [below left=1.8 cm of a] (i2) {$e^+$};
    
    \vertex [above right=2.1cm of b] (f1) {$\nu_\alpha$};
    \vertex [below right=2.1cm of b] (f2) {$N$};

    \diagram* {
      (i1) -- [fermion] (a),
      (i2) -- [anti fermion] (a),
      (a) -- [boson, edge label'=$Z$] (b),
      (b) -- [plain] (f1),
      (b) -- [plain] (f2),
    };

    \vertex [right=6cm of a] (x_in) {$N$}; 
    \vertex [right=2.4cm of x_in, dot, label=95:$U^2$] (v_dec) {};
    
    \vertex [above right=2.1cm of v_dec] (out1) {$ \nu_\beta, \ell_\beta, \ell_\beta$};
    \vertex [right=1.9cm of v_dec] (out3) {$f, \nu_\gamma, u$};
    \vertex [below right=2.1cm of v_dec] (out2) {$\bar{f}, \bar{\ell}_\gamma, \bar{d}$};
    
    \diagram* {
      (x_in) -- [plain] (v_dec),
      (v_dec) -- [plain] (out1),
      (v_dec) -- [plain] (out3),
      (v_dec) -- [plain] (out2), 
    };
  \end{feynman}
\end{tikzpicture}
    \caption{\emph{Left:}
    Feynman diagram representing the HNL production from the decay of an on-shell Z-boson. 
    \emph{Right}: Feynman diagram symbolically summarising the various HNL decay channels mediated by both the neutral current \eqref{HNLdecaysNC} and charged current \eqref{HNLdecaysCC}, here represented by a four-fermion interaction because the gauge boson is off-shell. }
    \label{fig:feynman-2}
\end{figure*}
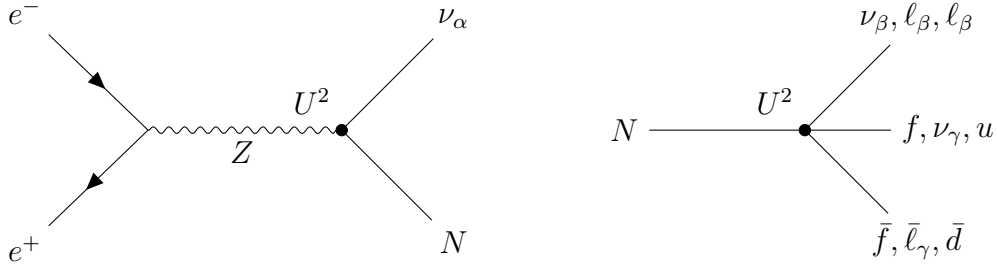

\begin{figure}
\centering
\includegraphics[width=0.45\textwidth]{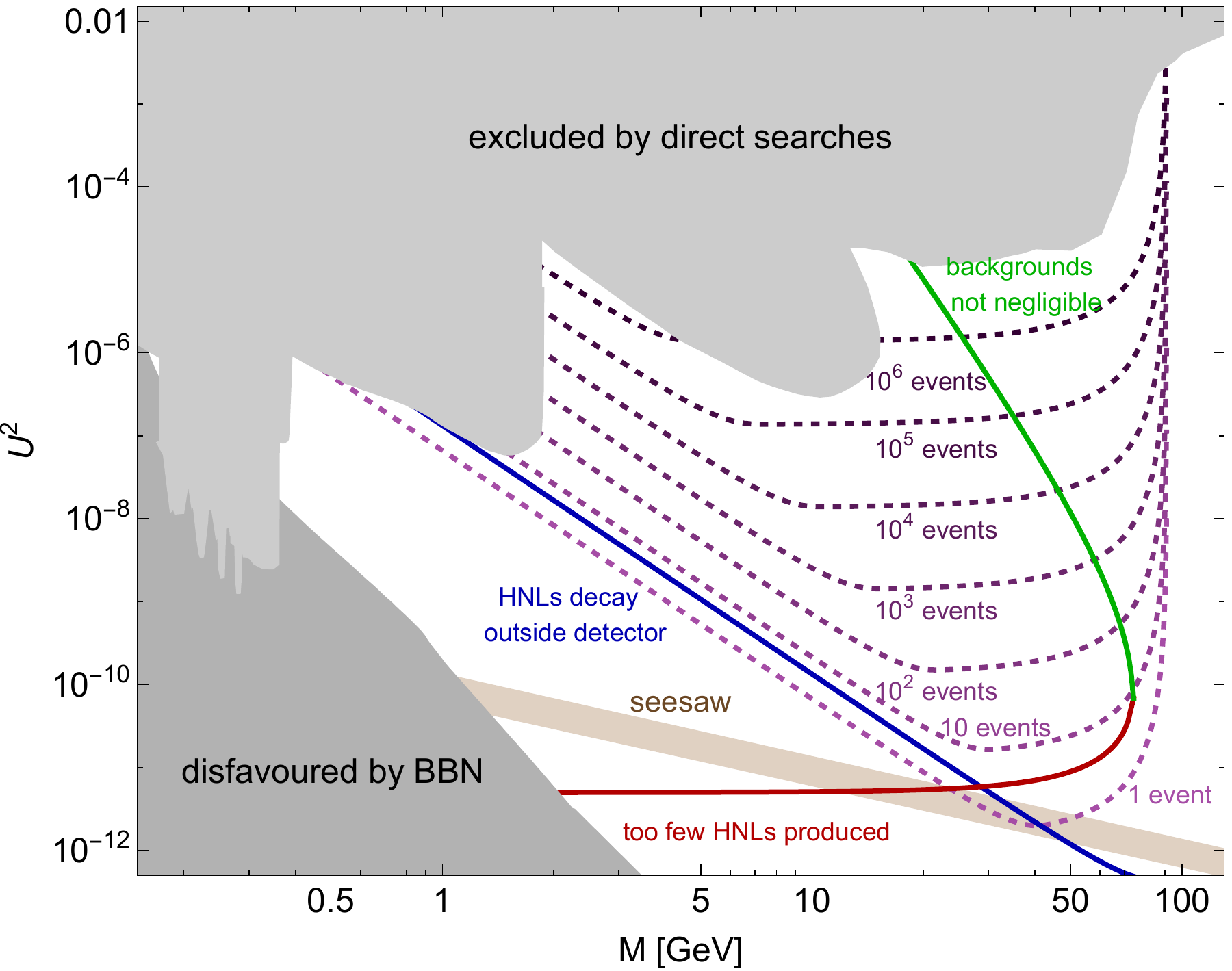} 
\caption{\label{fig:HNLdiscovery1}
The dashed lines represent the expected total number of HNL decays 
inside a cylindrical fiducial volume of diameter 
\MaD{$d_{\rm cyl} = 9$m} 
and length 
\MaD{$l_{\rm cyl} = 9$m }
with $N_Z = 6\times 10^{12}$ and $U^2=U_\mu^2$, as obtained from \eqref{NobsGeneral}. We set $l_0 = 0$ to emphasise the fact that any event inside the fiducial volume can in principle be observed. The green line is based on \eqref{Kenny2} and indicates the regime in which more than $\n=4$ events with displacements exceeding 
\MaD{$l_0 = 500 \, \mu$m \cite{Bellagamba:2025xpd}} 
are expected, a \new{longevity} cut which we assume to remove most SM backgrounds.\protect\footnotemark 
The red and blue lines indicate the corresponding limitations coming from the integrated luminosity and the detector dimensions for $\n=4$, based on \eqref{Kenny} and \eqref{Methusalix}, respectively. 
\MaD{Roughly speaking, the three lines indicate  frontiers that can be pushed by different strategies: more integrated luminosity (red), bigger detectors (blue), removing backgrounds (green).\\}
\MaD{For comparison,} the gray shaded areas represent the parameter regions excluded by experimental searches \cite{Abdullahi:2022jlv}  
and big bang nucleosynthesis \cite{Boyarsky:2020dzc}.
The brown band represents the  \emph{seesaw floor} \eqref{seesaw}, its width indicates the uncertainty in this lower bound when varying the sum of SM neutrino masses between the lowest possible value consistent with oscillation data \cite{Esteban:2024eli} and the upper bound from Planck observations \cite{Planck:2018vyg}. 
Heavy neutrinos with masses and mixings anywhere in the white region above this lower bound can simultaneously explain the light neutrino masses and the matter-antimatter asymmetry of the universe for technically natural parameter choices \cite{Drewes:2021nqr}.
A significant fraction of the wedge between the blue line, the seesaw floor and the BBN-disfavoured region can be explored with fixed target experiments or additional detectors at the LHC or future colliders \cite{Beacham:2019nyx,Agrawal:2021dbo}; the most sensitive approved experiments are the near detector of DUNE \cite{DUNE:2015lol,DUNE:2020ypp} and SHiP \cite{SHiP:2015vad,Alekhin:2015byh,SHiP:2025ows}. 
}
\end{figure}

\begin{figure}
\centering
\includegraphics[width=0.45\textwidth]{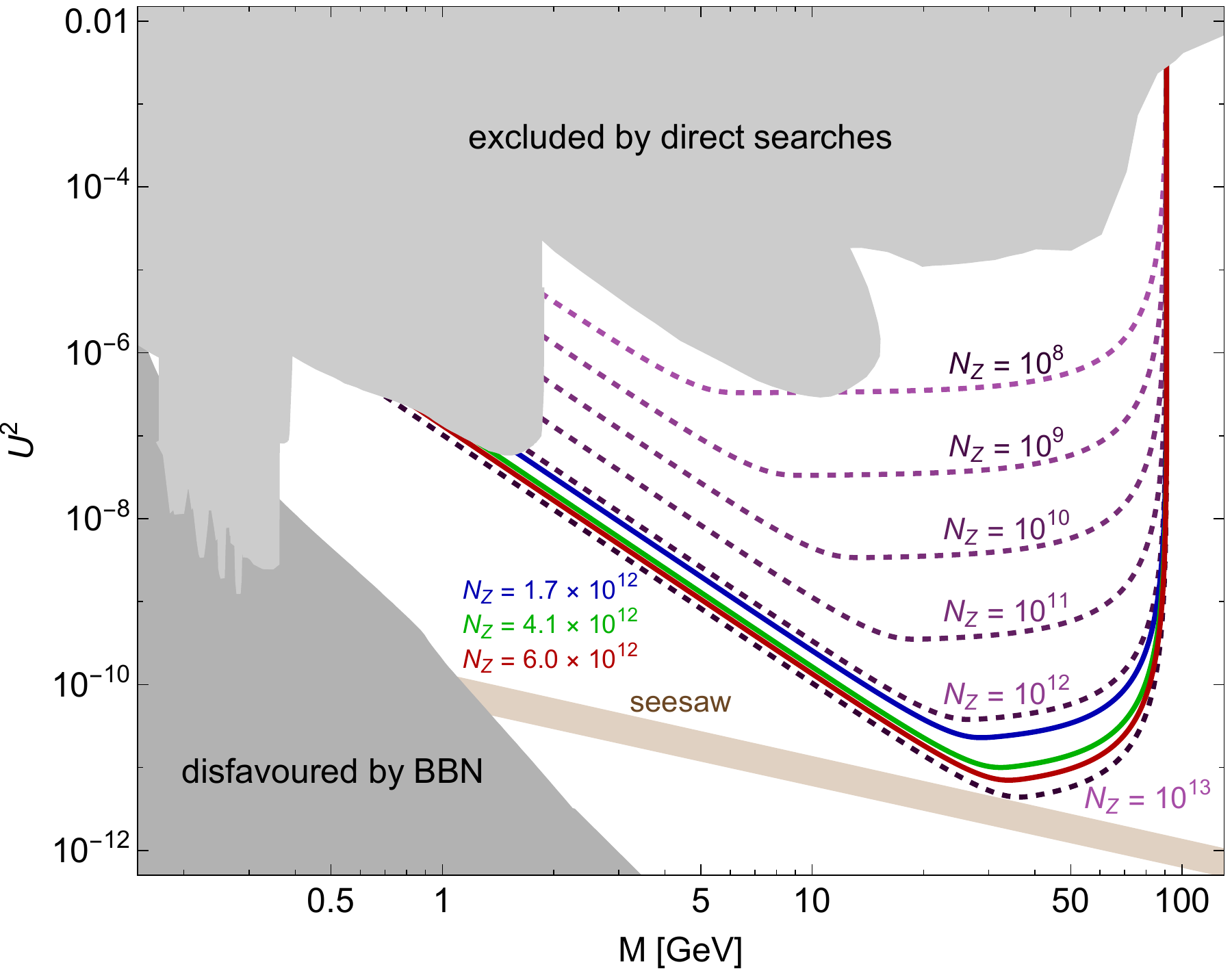} 
\caption{\label{fig:HNLdiscovery2}
The dashed lines indicate the integrated luminosity for which $\n=4$ HNL decays are expected within the fiducial volume, with all other parameter choices as in Fig.~\ref{fig:HNLdiscovery1}.
The colourful lines represent the reach that could be achieved with the numbers given in table \ref{tab:NZ}, reflecting the current planning for several proposed Tera-Z factories.
}
\end{figure}

We consider HNL production from s-channel Z-bosons with subsequent HNL decays via both neutral and charged currents, cf.~Fig.~\ref{fig:feynman-2},
\begin{subequations}
\begin{eqnarray}
e^+e^- &\to& Z \ \to \ \nu_\alpha N \ \to \ \nu_\alpha \ \nu_\beta f\bar{f}\label{HNLdecaysNC}\\
e^+e^-  &\to&  Z \ \to \ \nu_\alpha N \ \to \ \nu_\alpha \ \ell_\beta \nu_\gamma\bar{\ell}_\gamma
\ {\rm or} \ 
\nu_\alpha \ \ell_\beta u\bar{d}
. \label{HNLdecaysCC}
\end{eqnarray}
\end{subequations}
It is convenient to introduce the quantities
\begin{eqnarray}
    U_\alpha^2 = |\theta_\alpha|^2 \ , \ U^2 = \sum_\alpha U_\alpha^2.
\end{eqnarray}
Since HNL production and decay are both controlled by the mixing in the model \eqref{PhenoModelLagrandian} we can associate them with a single small parameter
\begin{eqnarray}
\upepsilon_{\rm pro} = \upepsilon_{\rm dec}  =  U^2 = \upepsilon.
\end{eqnarray}
For the energies relevant at Tera-Z factories the masses of all accessible final state particles can be neglected, then the HNL momentum is ${\rm p}_N=\frac{m_Z}{2}\left(1 - (M/m_Z)^2\right)$.
The HNL decay rate $\GammaN$ in \eqref{DecayLength} is
given by \eqref{GammaLLP} with \cite{Atre:2009rg}
\begin{eqnarray}\label{GammaDefinition}
\GammaNvis \simeq  \frac{11\cdec}{192\pi^3} M^5 G_F^2, \quad 
\GammaNinv \simeq  \frac{\cdec}{192\pi^3} U^2 M^5 G_F^2.
\end{eqnarray}
Here $\cdec = 2$ for Majorana and $\cdec = 1$ for Dirac HNLs. The phenomenology of realistic models with several HNL flavours may effectively be described by choosing non-integer values of $\cdec$.\footnote{Discussions of the mapping between the Lagrangian \eqref{PhenoModelLagrandian} and realistic neutrino mass models can e.g.~be found in \cite{Drewes:2022akb,Drewes:2024pad}.}
The numbers of observed events can then be obtained from \eqref{NobsGeneral} with \cite{Antusch:2015mia}
\begin{eqnarray}
&& \A = \frac{2}{15}
\left(\frac{2{\rm p}_N}{m_Z}\right)^2
\left(
1+\frac{(M/m_Z)^2}{2}
\right) \ , \\
&& {\rm Br}(\N\to  \YY_\beta) = \frac{\upepsilon_{\rm dec} \Bnbeta}{\GammaN} = \frac{U_\beta^2}{U^2}.
\end{eqnarray}
Figs.~\ref{fig:HNLdiscovery1} and \ref{fig:HNLdiscovery2} illustrate the discovery potential for HNLs with masses below the electroweak scale at Tera-Z factories. 
The sensitivity region does not only extend beyond the reach of the HL-LHC \cite{Drewes:2019fou} by more than an order of magnitude, but gets close to the so-called seesaw-line , i.e., the lower bound on $U^2$ from the requirement to explain the light neutrino masses $m_i$ (see e.g.~\cite{Drewes:2019mhg}),
\begin{eqnarray}\label{seesaw}
U^2 > \frac{\sum_i m_i}{M}.    
\end{eqnarray}

\begin{figure}
\centering
\includegraphics[width=0.45\textwidth]{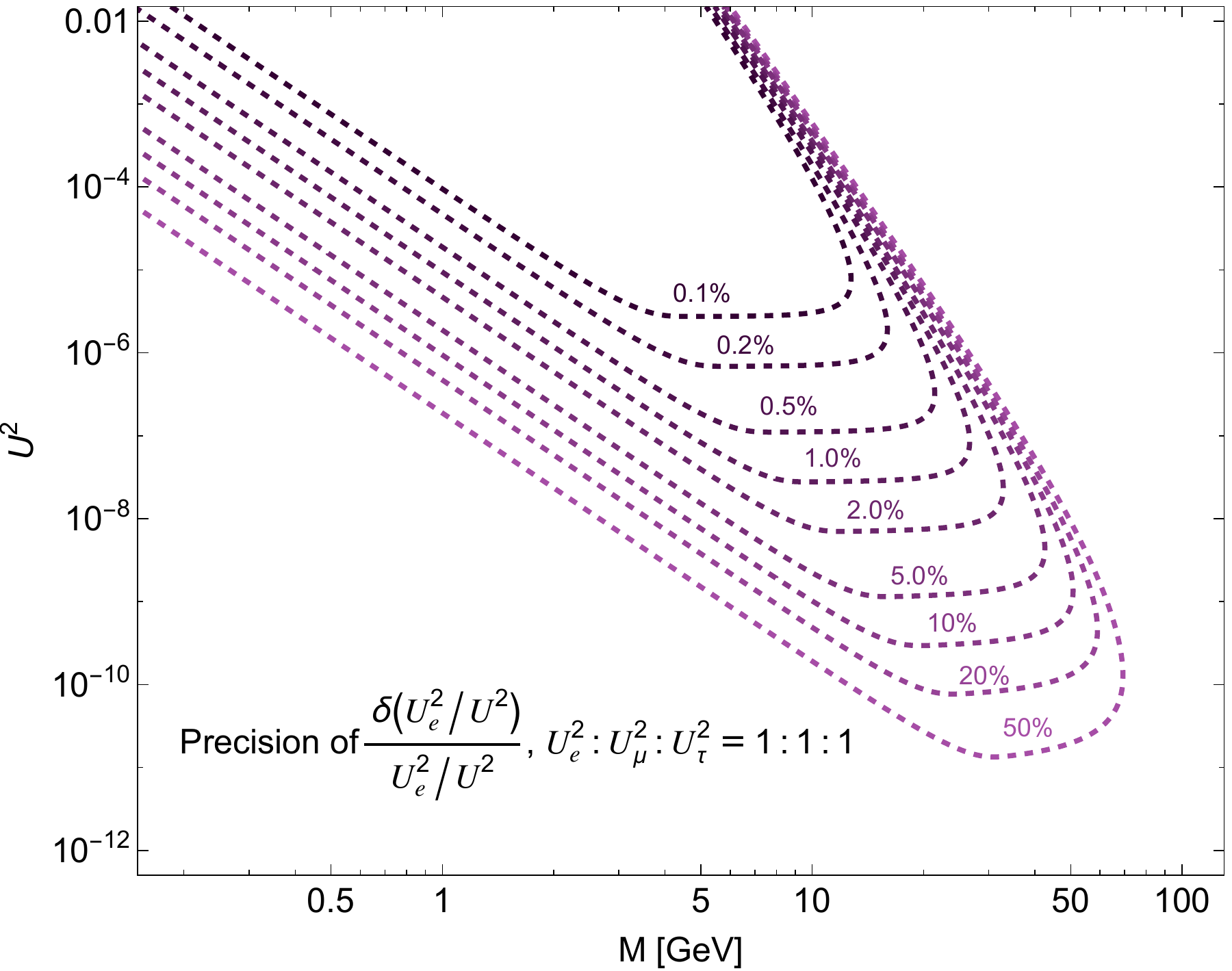} 
\includegraphics[width=0.45\textwidth]{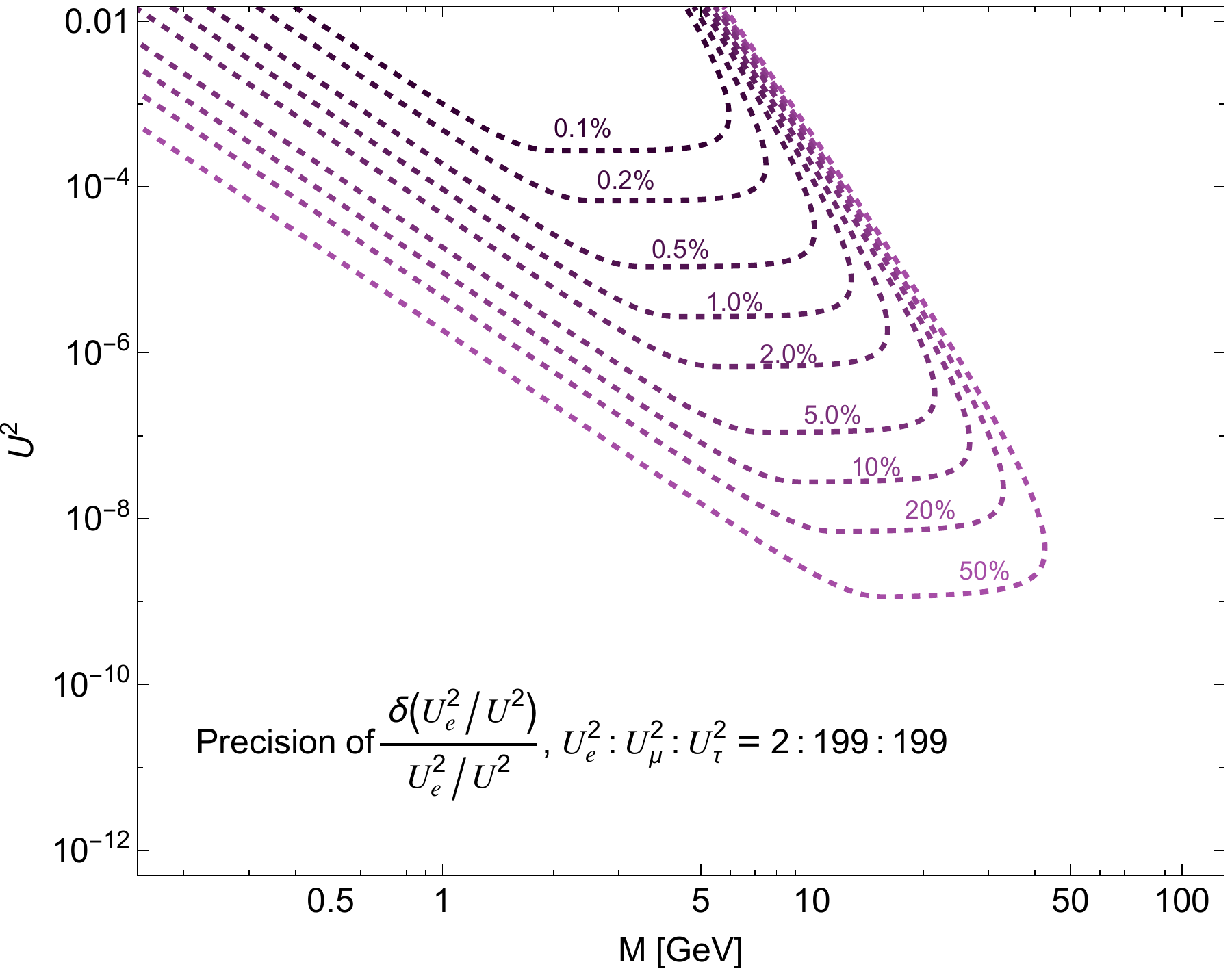} 
\caption{\label{fig:HNLprecision}
The dashed lines indicate the precision at which the branching ratios in HNL decays can be measured based on \eqref{HNLprecision} with 
\MaD{$l_0 = 500 \mu m$}
and all other parameters as in Fig.~\ref{fig:HNLdiscovery1}.
In the upper panel we assume a ratio $U_e^2:U_\mu^2:U_\tau^2 = 1:1:1$,
in the lower panel we assume $U_e^2:U_\mu^2:U_\tau^2 = 2 : 199 : 199$.  
The former is favoured for an inverted ordering of the light neutrino masses, the latter for a normal ordering, cf.~\cite{Drewes:2022akb} and references therein. 
}
\end{figure}

The large number of events in Fig.~\ref{fig:HNLdiscovery1} implies that Tera-Z factories can not only discover HNLs, but also study their properties in detail. 
One important observable is the potential lepton number violation in HNL decays, which has e.g.~been studied in \cite{Blondel:2021mss,Drewes:2022rsk,Antusch:2023jsa,Antusch:2024otj}.
Another important set of observables are the branching ratios for HNL decays into final states with individual SM flavours $\beta=e,\mu,\tau$.
If the masses of all final state particles are negligible, these are entirely fixed by the ratios $U_\beta^2/U^2$.
If HNLs are observed, measurements of these ratios could discriminate between models and shed light on their potential role in neutrino masses and leptogenesis.
The accuracy at which the $U_\beta^2/U^2$ will be measured can be estimated by means of \eqref{precision},
\begin{eqnarray}\label{HNLprecision}
    \frac{\delta(U_\beta^2/U^2)}{U_\beta^2/U^2} \simeq \sqrt{
\frac{1}{N_{\rm obs}}\left(U^2/U_\beta^2 - 1\right)},
\end{eqnarray}
with $N_{\rm obs}$ the total number of events obtained from \eqref{NobsGeneral} by summing over all observable final states.
Fig.~\ref{fig:HNLprecision} illustrates that Tera-Z factories could measure the HNL branching ratios at the percent level. 
In the $\nu$MSM
the $U_\beta^2/U^2$ only depend on light neutrino properties; measuring them would provide a powerful test of the seesaw mechanism and leptogenesis \cite{Hernandez:2016kel,Drewes:2016jae,Antusch:2017pkq,Hernandez:2022ivz}, in particular when combined with an observation of neutrinoless double $\beta$-decay \cite{Drewes:2016lqo,Hernandez:2016kel,deVries:2024rfh}. In less minimal models the $U_\beta^2/U^2$ also depend on additional parameters in the sterile sector. However, the large number of events expected at a Tera-Z factory still enable model discrimination and the measurements of several parameters \cite{Drewes:2024pad,Drewes:2024bla}.
This demonstrates that Tera-Z factories are powerful discovery and precision machines in one.

\footnotetext{\MaD{
Strictly speaking imposing $\lambdaN > 0.5$mm only removes the backgrounds when combined with other cuts, which would lead to a reduction of the signal efficiency 
\cite{Bellagamba:2025xpd}
which we ignore here.
}}

\subsection{Axion-like particles}

\begin{figure*}[ht]
    \centering
    \begin{tikzpicture}
  \begin{feynman}
    \vertex (a);                  
    \vertex [right=2.5cm of a, dot, label=95:$c_{\gamma Z}$] (b) {};
    
    \vertex [above left= 1.8 cm of a] (i1) {$e^-$};
    \vertex [below left=1.8 cm of a] (i2) {$e^+$};
    
    \vertex [above right=2.1cm of b] (f1) {$\gamma$};
    \vertex [below right=2.1cm of b] (f2) {$a$};

    \diagram* {
      (i1) -- [fermion] (a),
      (i2) -- [anti fermion] (a),
      (a) -- [boson, edge label'=$Z$] (b),
      (b) -- [boson] (f1),
      (b) -- [scalar] (f2),
    };

    \vertex [right=6cm of a] (x_in) {$a$}; 
    \vertex [right=3cm of x_in, dot, label=95:$c_{\gamma \gamma}$] (v_dec) {};
    
    \vertex [above right=2.1cm of v_dec] (out1) {$\gamma$};
    \vertex [below right=2.1cm of v_dec] (out2) {$\gamma $};
    
    \diagram* {
      (x_in) -- [scalar] (v_dec),
      (v_dec) -- [boson] (out1),
      (v_dec) -- [boson] (out2), 
    };
  \end{feynman}
\end{tikzpicture}
    \caption{Feynman diagrams representing the production of an ALP from an on-shell Z-boson and its subsequent decay into photons, cf.~\eqref{ALPdecay}.}
    \label{fig:feynman-3}
\end{figure*}
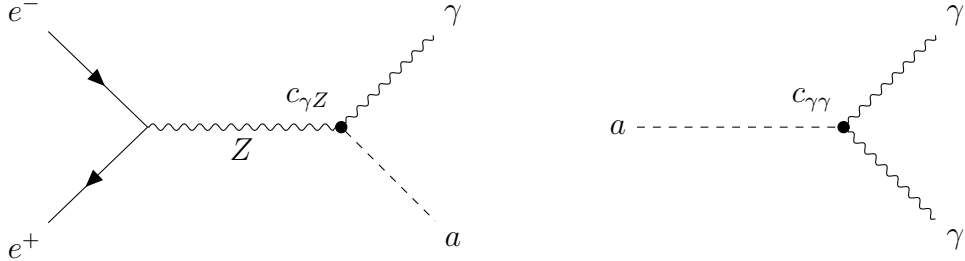

\begin{figure}
\centering
\includegraphics[width=0.45\textwidth]{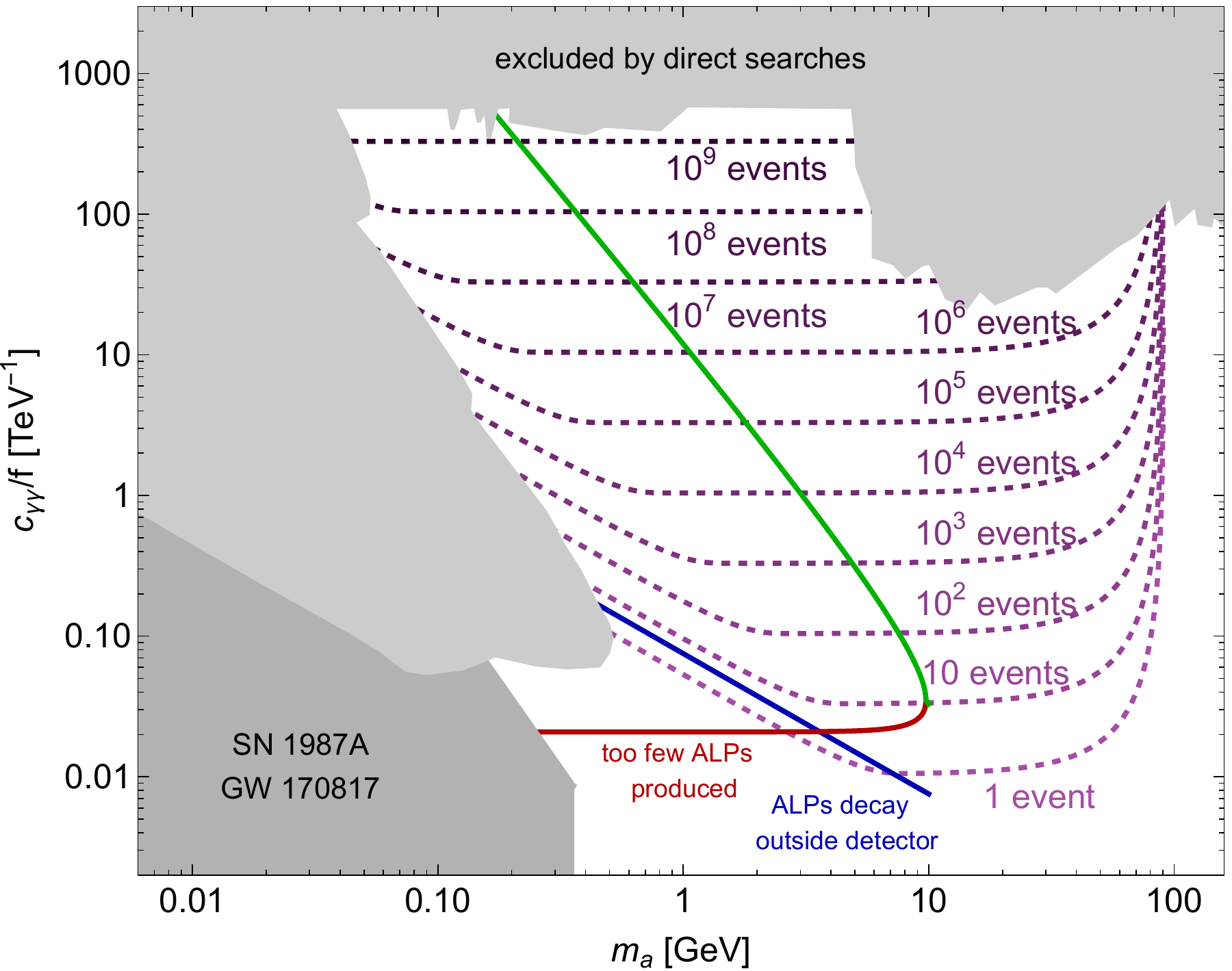} 
\includegraphics[width=0.45\textwidth]{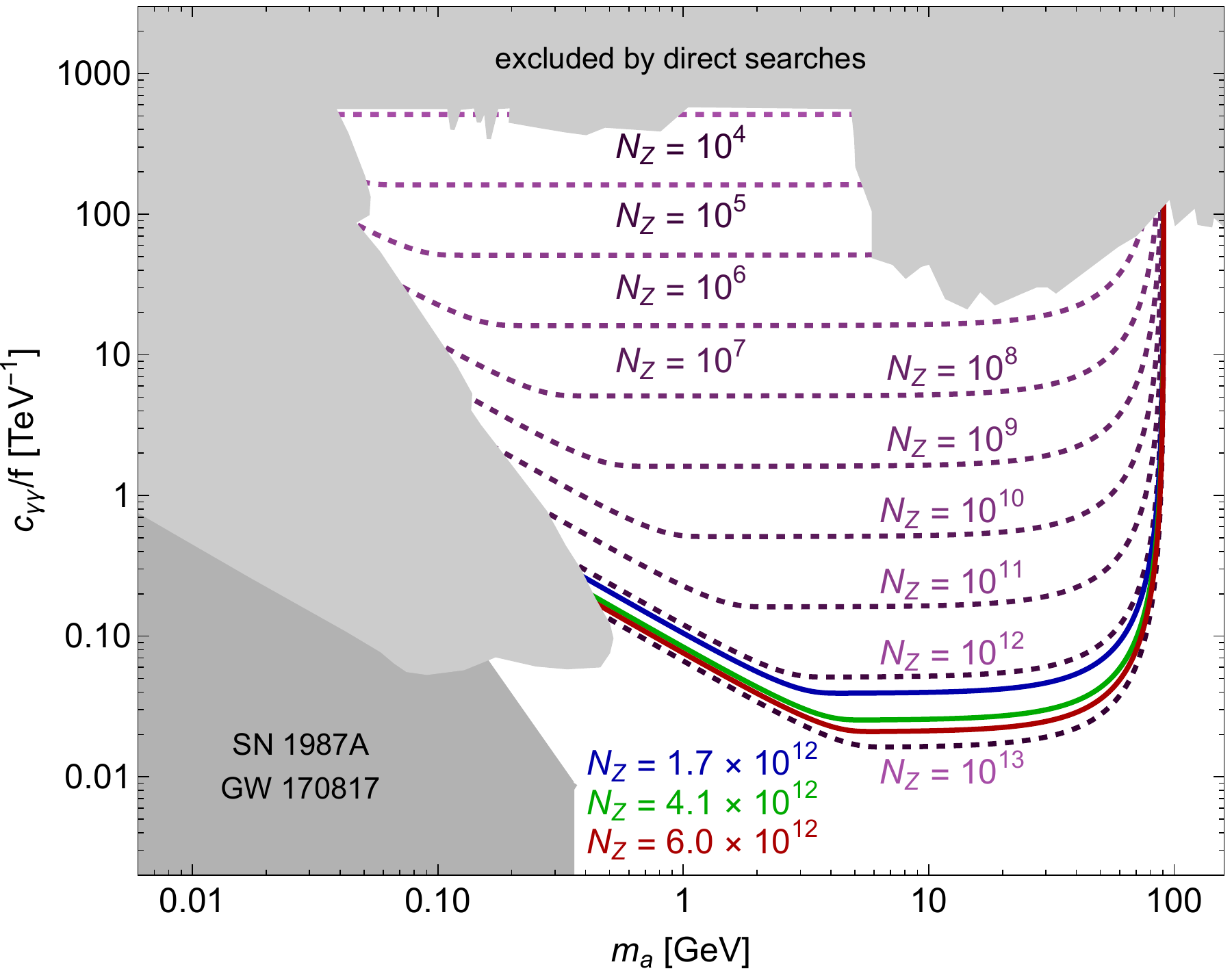} 
\caption{\label{fig:ALPs}
\emph{Upper panel:} 
The dashed lines represent the expected total number of ALP decays \eqref{ALPdecay} inside a cylindrical fiducial volume of diameter 
\MaD{$d_{\rm cyl}=8$m}
and length 
\MaD{$l_{\rm cyl}=8$m} 
with $N_Z = 6\times 10^{12}$, as obtained from \eqref{NobsGeneral}. 
The red and blue lines indicate the limitations coming from the integrated luminosity and the detector dimensions, based on \eqref{Kenny} and \eqref{Methusalix}, respectively, for $\n=4$. 
\new{The green line corresponds to $l_0=3$cm, roughly twice the resolution on the photon pointing variable measured by the ATLAS calorimeter \cite{ATLAS:2023meo}. }
The gray shaded areas represent the parameter regions excluded by experimental searches \cite{Antel:2023hkf}. 
\emph{Lower panel:}
The dashed lines indicate the integrated luminosity for which $\n=4$ ALP decays are expected within the fiducial volume for $N_Z$ as indicated in the plot. The colourful lines represent the reach that could be achieved with the numbers given in table \ref{tab:NZ}, reflecting the current planning for several proposed Tera-Z factories.
}
\end{figure}

Axions or ALPs $a$ with very light mass $m_a$ have initially been proposed in the context of the strong CP-problem~\cite{Peccei:1977ur,Peccei:1977hh,Wilczek:1977pj,Weinberg:1977ma}
and as Dark Matter candidates \cite{Preskill:1982cy,Dine:1982ah,Abbott:1982af} (c.f.~\cite{DiLuzio:2020wdo} and~\cite{Adams:2022pbo} for reviews, respectively).  They  could also drive cosmic inflation (cold \cite{Freese:1990rb,Kim:2004rp,Dimopoulos:2005ac,Silverstein:2008sg} or warm \cite{Berghaus:2019whh,Berghaus:2025dqi}).
However, in a more general context, also heavier ALPs 
with $m_a$ in the collider range 
have been considered  (see e.g.~\cite{Mimasu:2014nea,Bauer:2017ris}), including  long-lived ALPs at Tera-Z factories \cite{Bauer:2018uxu,RebelloTeles:2023uig,Bao:2025tqs,Polesello:2025gwj,Blondel:2022qqo,FCC:2025lpp,Altmann:2025feg}.
In general the ALPs couple to gauge bosons, the Higgs field and fermions through various dimension five operators. For the present purpose we consider a scenario where the relevant interactions can be described by the effective coupling to the photon and $Z$-boson field strength tensors $F^{\mu\nu}$ and $Z^{\mu\nu}$, 
\begin{eqnarray}\label{LAPL}
\mathcal{L}  &\supset&  e^2\,c_{\gamma\gamma}\,\frac{a}{\Lambda}\,F_{\mu\nu}\,\tilde F^{\mu\nu}
+ \frac{2e^2}{s_W c_W}\,c_{\gamma Z}\,\frac{a}{\Lambda}\,F_{\mu\nu}\,\tilde Z^{\mu\nu} \nonumber \\
&& + \frac{e^2}{s_W^2 c_W^2}\,c_{ZZ}\,\frac{a}{\Lambda}\,Z_{\mu\nu}\,\tilde Z^{\mu\nu} \,.
\end{eqnarray}
with $e$ the elementary charge, $s_W$ and $c_W$ the $\sin$ and $\cos$ of the Weinberg angle and $\Lambda =  (4\pi)^2 f$ a new physics scale that is often expressed in terms of an axion decay constant $f$.
The relations between the Wilson coefficients $c_{\ldots}$ depend on the UV completion of the effective model \eqref{LAPL}.
For searches at the Z-pole, we consider the process shown in Fig.~\ref{fig:feynman-3},
\begin{eqnarray}\label{ALPdecay}
    e^+e^-\to Z \to a \ \gamma \to \gamma\gamma \ \gamma.
\end{eqnarray}
The decay rate of Z-bosons into ALPs is \cite{Bauer:2018uxu}
\begin{eqnarray}
    \Gamma(Z\to \gamma a) = \frac{\alpha^2 m_Z^3}{96 \pi^3 s_W^2 c_W^2 f^2} |c_{\gamma Z}|^2
    \left( 1 - \frac{m_a^2}{m_Z^2} \right)^3,
\end{eqnarray}
and the subsequent decay rate of the axion is
\begin{eqnarray}
     \Gamma(a\to\gamma\gamma)
   = \frac{\alpha^2\space m_a^3}{64\pi^3 }  \frac{c_{\gamma\gamma}^2}{f^2}
\end{eqnarray}
with $\alpha = e^2/(4\pi)$.
Since the photon is massless, ALP momentum entering the decay length $\lambda_a$ in \eqref{DecayLength} is 
\begin{equation}
    {\rm p}_a=\frac{m_Z}{2}\left(1 - (m_a/m_Z)^2\right)
\end{equation}
In terms of the other quantities defined in section \ref{sec:GenericLLPanalytic} this corresponds to 
\begin{eqnarray}
&&\upepsilon_{\rm dec} = c_{\gamma\gamma}^2\frac{m_Z^2}{f^2} \ , \ 
\GammaNvis = \frac{\alpha^2\space m_a^3}{64\pi^3 m_Z^2} \ , \\
&&\upepsilon_{\rm pro} = \frac{c_{\gamma Z}^2}{s_W^4}\frac{m_Z^2}{f^2} \ , \
\A = \frac{m_Z}{\Gamma_Z}
\frac{s_W^2}{c_W^2}
\frac{\alpha^2}{96 \pi^3 } \left( 1 - \frac{m_a^2}{m_Z^2} \right)^3.
\end{eqnarray}
As a benchmark we consider the choice  $c_{\gamma Z} = -s_W^2 c_{\gamma \gamma}$, corresponding to a situation where the ALP only couples to the $U(1)$ hypercharge gauge boson before electroweak symmetry breaking, implying that the production and decay are governed by a single small parameter 
\begin{eqnarray}
\upepsilon_{\rm dec} = \upepsilon_{\rm pro} = c_{\gamma\gamma}^2\frac{m_Z^2}{f^2} =     \upepsilon.
\end{eqnarray}
With this at hand, we can immediately compute the number of events from \eqref{NobsGeneral}. The resulting event numbers are shown in Fig.~\ref{fig:ALPs}.

\section{Discussion and conclusion}
Tera-Z factories are versatile multi-purpose machines that can probe fundamental physics in numerous ways.
In addition to precise measurements of the Higgs boson and top quark properties, electroweak precision tests and flavour observables, they also have the potential to discover new elementary particles. A particularly promising avenue is searches for long-lived particles.  

In section \ref{sec:GenericLLPanalytic} we used simple analytic estimates to identify the sensitivity regions of long-lived particle searches at the Z-pole as a function of the integrated luminosity and the fiducial detector volume. The former is characterised by the number of produced Z-bosons $N_Z$, the latter are parameterised by a characteristic length scale $l_1$ that is related to the dimensions of a given detector by \eqref{l1def}.
Here \emph{sensitivity} is defined by the requirement that a  minimal number of events $\n$ are expected to be observed within the fiducial volume. The sensitivity region is  expressed in terms of the smallest and largest couplings  
fulfilling this requirement 
for a given particle mass $\M$. To be entirely model independent, we expressed the LLP couplings in terms of generic small parameters $\upepsilon_{\rm pro}$ and $\upepsilon_{\rm dec}$ associated with the production and decay of the new particles, respectively.

In section \ref{sec:examples} we applied the analytic formulae from section \ref{sec:GenericLLPanalytic} to two well-motivated models, namely HNLs and ALPs.  For HNLs, $\upepsilon_{\rm pro}$ and $\upepsilon_{\rm dec}$ can be related to their mixing with ordinary neutrinos, $\upepsilon_{\rm pro} = \upepsilon_{\rm dec} = U^2$. For the ALP model under consideration they express the separation of scales between their decay constant $f$ and the electroweak scale, $\upepsilon_{\rm pro} = \upepsilon_{\rm dec} = c_{\gamma\gamma}^2 (m_Z/f)^2$. Figures \ref{fig:HNLdiscovery2} and \ref{fig:ALPs} demonstrate that the sensitivity regions of Tera-Z factories 
with luminosities given in table \ref{tab:NZ}
extend beyond the current bounds by orders of magnitude in these small parameters. Moreover, deep within this region millions of events could be observed, enabling detailed studies of the properties of the discovered particles. We illustrate this by showing the precision by which HNL branching ratios can be measured in Fig.~\ref{fig:HNLprecision}.
This demonstrates that, in the context of searches for new particles with masses below the electroweak scale, Tera-Z factories are powerful discovery and precision machines in one. The remaining low mass region can be explored with the fixed target experiments such as SHiP or dedicated far detectors. 

Of course, the accuracy of our estimates falls short of current state-of-the-art simulations, and our analytic formulae can by no means replace them.
 However, comparing them to more realistic simulations \cite{Ajmal:2024kwi,Bellagamba:2025xpd,Bauer:2018uxu,Bao:2025tqs,Antusch:2024otj,Polesello:2025gwj} reveals that their accuracy goes beyond order of magnitude estimates.\footnote{The precise level of accuracy depends on the specific search under consideration, but generally the mismatch in the predicted number of observed events amounts to factors of order one. This makes little difference in double-logarithmic sensitivity plots like Figs.~\ref{fig:HNLdiscovery1}, \ref{fig:HNLdiscovery2}, \ref{fig:ALPs}.} We expect the same to hold for other LLPs.
This can be understood for two reasons.
Firstly, searches for displaced signatures from LLPs are approximately background-free, and realistic reconstruction efficiencies typically only affect the number of observed events $N_{\rm obs}$ by factors of order one (or at most factors of a few). At the same time $N_{\rm obs}$ in \eqref{NobsGeneral} exhibits a steep  dependence on the small couplings encoded in $\upepsilon_{\rm pro}$ and $ \upepsilon_{\rm dec}$, the allowed range of which typically covers several orders of magnitude. Hence, varying $N_{\rm obs}$ by a factor of order one affects the extension of the sensitivity region only mildly. 
Secondly, the Z-bosons considered here decay at rest. The momenta
$\pN$ of LLPs produced in $1\to 2$ decays are monochromatic and the phase space is fixed by their mass $\M$ alone. This is in contrast to the LHC or fixed target experiments, where the need to average over the spectra considerably reduces the accuracy of analytic estimates for $N_{\rm obs}$ and the resulting sensitivity regions \cite{Drewes:2019vjy,Bondarenko:2019yob}.

In view of this, our results are, in spite of the fact that more accurate sensitivity studies already exist in the literature, useful in two ways.
Firstly, they can be regarded as hard theoretical limits of what can be achieved under idealised conditions. This can help to  gain an intuition for the sensitivity gain that can be achieved by improving on those aspects that we have idealised (reconstruction efficiencies, triggers, backgrounds etc.). 
Secondly, the analytic approach makes the parametric dependence of the sensitivity on various limiting factors explicit (e.g.~integrated luminosity $N_Z$,  detector dimensions $l_1$, cuts on the displacement $l_0$).  This scaling can be used to understand which modification of a given machine or detector design leads to the biggest improvements. Moreover, with the analytic formulae, plots like figures \ref{fig:HNLdiscovery1}-\ref{fig:ALPs} can be produced within seconds for arbitrary parameter choices. This is useful to roughly chart the landscape of potential designs before studying the most promising ones with more accurate simulations. The code to generate the sensitivity curves and precision estimates from this work is publicly available at \href{https://github.com/liyuanzhen98/LLPatTeraZ}{\faGithub\ LLPatTeraZ}, it can easily be extended to other LLP scenarios.


\section*{Acknowledgements}

MaD would like to thank Elena Shaposhnikova and Mikhail Shaposhnikov for countless fruitful discussions and valuable input, and Jim Virdee for his insights on LEP3. \MaD{MaD is also very grateful to Giacomo Polesello for the detailed explanations on experimental aspects. }
A special thanks goes to
Alain Blondel, 
Admir Greljo,
Patrick Janot,
Lingfeng Li and
Guy Wilkinson for patiently answering my naive questions about precision tests.
YL would like to thank Yang Ma for the discussion of current constraints on ALPs.
This work has been partially funded by the National Key R\&D Program of China No. 2020YFC2201601 and by the Deutsche Forschungsgemeinschaft (DFG, German Research Foundation) - SFB 1258 - 283604770.

\bibliographystyle{JHEP}
\bibliography{bib_new.bib}{}

@article{ATLAS:2023meo,
    author = "Aad, Georges and others",
    collaboration = "ATLAS",
    title = "{Search in diphoton and dielectron final states for displaced production of Higgs or Z bosons with the ATLAS detector in s=13{\,}{\,}TeV pp collisions}",
    eprint = "2304.12885",
    archivePrefix = "arXiv",
    primaryClass = "hep-ex",
    reportNumber = "CERN-EP-2023-044",
    doi = "10.1103/PhysRevD.108.012012",
    journal = "Phys. Rev. D",
    volume = "108",
    number = "1",
    pages = "012012",
    year = "2023"
}

@article{Abbott:1982af,
 author = {Abbott, L. F. and Sikivie, P.},
 doi = {10.1016/0370-2693(83)90638-X},
 editor = {Srednicki, M. A.},
 journal = {Phys. Lett. B},
 pages = {133--136},
 reportnumber = {PRINT-82-0695 (BRANDEIS)},
 title = {{A Cosmological Bound on the Invisible Axion}},
 volume = {120},
 year = {1983}
}

@article{Abdullahi:2022jlv,
 archiveprefix = {arXiv},
 author = {Abdullahi, Asli M. and others},
 doi = {10.1088/1361-6471/ac98f9},
 eprint = {2203.08039},
 journal = {J. Phys. G},
 number = {2},
 pages = {020501},
 primaryclass = {hep-ph},
 reportnumber = {FERMILAB-CONF-22-184-T-V},
 title = {{The present and future status of heavy neutral leptons}},
 volume = {50},
 year = {2023}
}

@inproceedings{Adams:2022pbo,
 archiveprefix = {arXiv},
 author = {Adams, C. B. and others},
 booktitle = {{Snowmass 2021}},
 eprint = {2203.14923},
 month = {3},
 note = {arXiv: \href{https://arxiv.org/abs/2203.14923}{2203.14923} [hep-ex]},
 primaryclass = {hep-ex},
 reportnumber = {FERMILAB-CONF-22-996-PPD-T},
 title = {{Axion Dark Matter}},
 year = {2022}
}

@article{Agrawal:2021dbo,
 archiveprefix = {arXiv},
 author = {Agrawal, Prateek and others},
 doi = {10.1140/epjc/s10052-021-09703-7},
 eprint = {2102.12143},
 journal = {Eur. Phys. J. C},
 number = {11},
 pages = {1015},
 primaryclass = {hep-ph},
 title = {{Feebly-interacting particles: FIPs 2020 workshop report}},
 volume = {81},
 year = {2021}
}

@article{Ai:2024nmn,
 archiveprefix = {arXiv},
 author = {Ai, Xiaocong and others},
 eprint = {2412.19743},
 month = {12},
 note = {arXiv: \href{https://arxiv.org/abs/2412.19743}{2412.19743} [hep-ex]},
 primaryclass = {hep-ex},
 title = {{Flavor Physics at CEPC: a General Perspective}},
 year = {2024}
}

@article{Ai:2025cpj,
 archiveprefix = {arXiv},
 author = {Ai, Xiaocong and others},
 doi = {10.1088/1674-1137/ae1194},
 eprint = {2505.24810},
 month = {5},
 primaryclass = {hep-ex},
 title = {{New Physics Search at the CEPC: a General Perspective}},
 year = {2025}
}

@article{Ajmal:2024kwi,
 archiveprefix = {arXiv},
 author = {Ajmal, Sehar and Azzi, Patrizia and Giappichini, Sofia and Klute, Markus and Panella, Orlando and Presilla, Matteo and Zuo, Xunwu},
 doi = {10.1007/JHEP05(2025)054},
 eprint = {2410.03615},
 journal = {JHEP},
 pages = {054},
 primaryclass = {hep-ph},
 title = {{Searching for type I seesaw mechanism in a two heavy neutral leptons scenario at FCC-ee}},
 volume = {05},
 year = {2025}
}

@article{Alekhin:2015byh,
 archiveprefix = {arXiv},
 author = {Alekhin, Sergey and others},
 doi = {10.1088/0034-4885/79/12/124201},
 eprint = {1504.04855},
 journal = {Rept. Prog. Phys.},
 number = {12},
 pages = {124201},
 primaryclass = {hep-ph},
 reportnumber = {CERN-SPSC-2015-017, SPSC-P-350-ADD-1},
 title = {{A facility to Search for Hidden Particles at the CERN SPS: the SHiP physics case}},
 volume = {79},
 year = {2016}
}

@article{Alimena:2019zri,
 archiveprefix = {arXiv},
 author = {Alimena, Juliette and others},
 doi = {10.1088/1361-6471/ab4574},
 eprint = {1903.04497},
 journal = {J. Phys. G},
 number = {9},
 pages = {090501},
 primaryclass = {hep-ex},
 title = {{Searching for long-lived particles beyond the Standard Model at the Large Hadron Collider}},
 volume = {47},
 year = {2020}
}

@book{Altmann:2025feg,
 archiveprefix = {arXiv},
 author = {Altmann, J. and others},
 doi = {10.23731/CYRM-2025-005},
 eprint = {2506.15390},
 isbn = {978-92-9083-700-8, 978-92-9083-701-5},
 month = {6},
 primaryclass = {hep-ex},
 reportnumber = {CERN-2025-005},
 series = {CERN Yellow Reports: Monographs},
 title = {{ECFA Higgs, electroweak, and top Factory Study}},
 volume = {5/2025},
 year = {2025}
}

@article{Anastopoulos:2025jyh,
 archiveprefix = {arXiv},
 author = {Anastopoulos, C. and others},
 eprint = {2504.00541},
 month = {4},
 note = {arXiv: \href{https://arxiv.org/abs/2504.00541}{2504.00541} [physics.acc-ph]},
 primaryclass = {physics.acc-ph},
 title = {{LEP3: A High-Luminosity e+e- Higgs and ElectroweakFactory in the LHC Tunnel}},
 year = {2025}
}

@article{Antel:2023hkf,
 archiveprefix = {arXiv},
 author = {Antel, C. and others},
 doi = {10.1140/epjc/s10052-023-12168-5},
 eprint = {2305.01715},
 journal = {Eur. Phys. J. C},
 number = {12},
 pages = {1122},
 primaryclass = {hep-ph},
 reportnumber = {CERN-TH-2023-061, DESY-23-050, FERMILAB-PUB-23-149-PPD, INFN-23-14-LNF, JLAB-PHY-23-3789, LA-UR-23-21432, MITP-23-015},
 title = {{Feebly-interacting particles: FIPs 2022 Workshop Report}},
 volume = {83},
 year = {2023}
}

@article{Antusch:2015mia,
 archiveprefix = {arXiv},
 author = {Antusch, Stefan and Fischer, Oliver},
 doi = {10.1007/JHEP05(2015)053},
 eprint = {1502.05915},
 journal = {JHEP},
 pages = {053},
 primaryclass = {hep-ph},
 reportnumber = {MPP-2015-24},
 title = {{Testing sterile neutrino extensions of the Standard Model at future lepton colliders}},
 volume = {05},
 year = {2015}
}

@article{Antusch:2016ejd,
 archiveprefix = {arXiv},
 author = {Antusch, Stefan and Cazzato, Eros and Fischer, Oliver},
 doi = {10.1142/S0217751X17500786},
 eprint = {1612.02728},
 journal = {Int. J. Mod. Phys. A},
 number = {14},
 pages = {1750078},
 primaryclass = {hep-ph},
 title = {{Sterile neutrino searches at future $e^-e^+$, $pp$, and $e^-p$ colliders}},
 volume = {32},
 year = {2017}
}

@article{Antusch:2016vyf,
 archiveprefix = {arXiv},
 author = {Antusch, Stefan and Cazzato, Eros and Fischer, Oliver},
 doi = {10.1007/JHEP12(2016)007},
 eprint = {1604.02420},
 journal = {JHEP},
 pages = {007},
 primaryclass = {hep-ph},
 title = {{Displaced vertex searches for sterile neutrinos at future lepton colliders}},
 volume = {12},
 year = {2016}
}

@article{Antusch:2017pkq,
 archiveprefix = {arXiv},
 author = {Antusch, Stefan and Cazzato, Eros and Drewes, Marco and Fischer, Oliver and Garbrecht, Bjorn and Gueter, Dario and Klaric, Juraj},
 doi = {10.1007/JHEP09(2018)124},
 eprint = {1710.03744},
 journal = {JHEP},
 pages = {124},
 primaryclass = {hep-ph},
 reportnumber = {TUM-1160/18, CP3-17-48},
 title = {{Probing Leptogenesis at Future Colliders}},
 volume = {09},
 year = {2018}
}

@article{Antusch:2023jsa,
 archiveprefix = {arXiv},
 author = {Antusch, Stefan and Hajer, Jan and Oliveira, Bruno M. S.},
 doi = {10.1007/JHEP10(2023)129},
 eprint = {2308.07297},
 journal = {JHEP},
 pages = {129},
 primaryclass = {hep-ph},
 title = {{Heavy neutrino-antineutrino oscillations at the FCC-ee}},
 volume = {10},
 year = {2023}
}

@article{Antusch:2023nqd,
 archiveprefix = {arXiv},
 author = {Antusch, Stefan and Hajer, Jan and Rosskopp, Johannes},
 doi = {10.1007/JHEP11(2023)235},
 eprint = {2307.06208},
 journal = {JHEP},
 pages = {235},
 primaryclass = {hep-ph},
 title = {{Decoherence effects on lepton number violation from heavy neutrino-antineutrino oscillations}},
 volume = {11},
 year = {2023}
}

@article{Antusch:2024otj,
 archiveprefix = {arXiv},
 author = {Antusch, Stefan and Hajer, Jan and Oliveira, Bruno M. S.},
 doi = {10.1007/JHEP11(2024)102},
 eprint = {2408.01389},
 journal = {JHEP},
 pages = {102},
 primaryclass = {hep-ph},
 title = {{Discovering heavy neutrino-antineutrino oscillations at the Z-pole}},
 volume = {11},
 year = {2024}
}

@techreport{Arduini:2947728,
 address = {Geneva},
 author = {Arduini, Gianluigi and Bordry, Frederick and Brinkmann,
Reinhard and Burrows, Philip and Desch, Klaus and
Farrington, Sinead and Gianotti, Fabiola and Hanagaki,
Kazunori and Holtkamp, Norbert and Keintzel, Jacqueline and
Kilminster, Ben and Lesiak, Tadeusz and Rivkin, Lenny and
Sabatié, Franck and Tuts, Mike and Zoccoli, Antonio},
 reportnumber = {CERN-ESU-ESG-WG2a-full-report},
 title = {{Assessment of large-scale accelerator projects at CERN -
Report of ESG WG2a}},
 url = {https://cds.cern.ch/record/2947728},
 year = {2025}
}

@article{Asaka:2005an,
 archiveprefix = {arXiv},
 author = {Asaka, Takehiko and Blanchet, Steve and Shaposhnikov, Mikhail},
 doi = {10.1016/j.physletb.2005.09.070},
 eprint = {hep-ph/0503065},
 journal = {Phys. Lett. B},
 pages = {151--156},
 title = {{The nuMSM, dark matter and neutrino masses}},
 volume = {631},
 year = {2005}
}

@article{Asaka:2005pn,
 archiveprefix = {arXiv},
 author = {Asaka, Takehiko and Shaposhnikov, Mikhail},
 doi = {10.1016/j.physletb.2005.06.020},
 eprint = {hep-ph/0505013},
 journal = {Phys. Lett. B},
 pages = {17--26},
 title = {{The $\nu$MSM, dark matter and baryon asymmetry of the universe}},
 volume = {620},
 year = {2005}
}

@article{Atre:2009rg,
 archiveprefix = {arXiv},
 author = {Atre, Anupama and Han, Tao and Pascoli, Silvia and Zhang, Bin},
 doi = {10.1088/1126-6708/2009/05/030},
 eprint = {0901.3589},
 journal = {JHEP},
 pages = {030},
 primaryclass = {hep-ph},
 reportnumber = {FERMILAB-PUB-08-086-T, NSF-KITP-08-54, MADPH-06-1466, DCPT-07-198, IPPP-07-99},
 title = {{The Search for Heavy Majorana Neutrinos}},
 volume = {05},
 year = {2009}
}

@article{Bao:2025tqs,
 archiveprefix = {arXiv},
 author = {Bao, Shou-shan and Ma, Yang and Wu, Yongcheng and Xie, Keping and Zhang, Hong},
 doi = {10.1007/JHEP10(2025)122},
 eprint = {2505.10023},
 month = {5},
 primaryclass = {hep-ph},
 reportnumber = {COMETA-2025-02, IRMP-CP3-25-09, MSUHEP-25-002, CPTNP-2025-011},
 title = {{Light Axion-Like Particles at Future Lepton Colliders}},
 year = {2025}
}

@article{Barducci:2020icf,
 archiveprefix = {arXiv},
 author = {Barducci, Daniele and Bertuzzo, Enrico and Caputo, Andrea and Hernandez, Pilar and Mele, Barbara},
 doi = {10.1007/JHEP03(2021)117},
 eprint = {2011.04725},
 journal = {JHEP},
 pages = {117},
 primaryclass = {hep-ph},
 title = {{The see-saw portal at future Higgs Factories}},
 volume = {03},
 year = {2021}
}

@article{Bauer:2017ris,
 archiveprefix = {arXiv},
 author = {Bauer, Martin and Neubert, Matthias and Thamm, Andrea},
 doi = {10.1007/JHEP12(2017)044},
 eprint = {1708.00443},
 journal = {JHEP},
 pages = {044},
 primaryclass = {hep-ph},
 reportnumber = {MITP-17-047},
 title = {{Collider Probes of Axion-Like Particles}},
 volume = {12},
 year = {2017}
}

@article{Bauer:2018uxu,
 archiveprefix = {arXiv},
 author = {Bauer, Martin and Heiles, Mathias and Neubert, Matthias and Thamm, Andrea},
 doi = {10.1140/epjc/s10052-019-6587-9},
 eprint = {1808.10323},
 journal = {Eur. Phys. J. C},
 number = {1},
 pages = {74},
 primaryclass = {hep-ph},
 reportnumber = {CERN-TH-2018-199, MITP/18-075},
 title = {{Axion-Like Particles at Future Colliders}},
 volume = {79},
 year = {2019}
}

@article{Beacham:2019nyx,
 archiveprefix = {arXiv},
 author = {Beacham, J. and others},
 doi = {10.1088/1361-6471/ab4cd2},
 eprint = {1901.09966},
 journal = {J. Phys. G},
 number = {1},
 pages = {010501},
 primaryclass = {hep-ex},
 reportnumber = {CERN-PBC-REPORT-2018-007},
 title = {{Physics Beyond Colliders at CERN: Beyond the Standard Model Working Group Report}},
 volume = {47},
 year = {2020}
}

@article{Bellagamba:2025xpd,
 archiveprefix = {arXiv},
 author = {Bellagamba, L. and Polesello, G. and Valle, N.},
 doi = {10.1140/epjc/s10052-025-14749-y},
 eprint = {2503.19464},
 journal = {Eur. Phys. J. C},
 number = {9},
 pages = {1069},
 primaryclass = {hep-ex},
 title = {{Searches for heavy neutral leptons at FCC-ee in final states including a muon}},
 volume = {85},
 year = {2025}
}

@article{Berghaus:2019whh,
 archiveprefix = {arXiv},
 author = {Berghaus, Kim V. and Graham, Peter W. and Kaplan, David E.},
 doi = {10.1088/1475-7516/2020/03/034},
 eprint = {1910.07525},
 journal = {JCAP},
 note = {[Erratum: JCAP 10, E02 (2023)]},
 pages = {034},
 primaryclass = {hep-ph},
 title = {{Minimal Warm Inflation}},
 volume = {03},
 year = {2020}
}

@article{Berghaus:2025dqi,
 archiveprefix = {arXiv},
 author = {Berghaus, Kim V. and Drewes, Marco and Zell, Sebastian},
 doi = {10.1103/9nn9-bsm9},
 eprint = {2503.18829},
 journal = {Phys. Rev. Lett.},
 number = {17},
 pages = {171002},
 primaryclass = {hep-ph},
 title = {{Warm Inflation with the Standard Model}},
 volume = {135},
 year = {2025}
}

@article{Blondel:2014bra,
 archiveprefix = {arXiv},
 author = {Blondel, Alain and Graverini, E. and Serra, N. and Shaposhnikov, M.},
 collaboration = {FCC-ee study Team},
 doi = {10.1016/j.nuclphysbps.2015.09.304},
 editor = {Aguilar-Ben{\'\i}tez, M and Fuster, J and Mart{\'\i}-Garc{\'\i}a, S and Santamar{\'\i}a, A},
 eprint = {1411.5230},
 journal = {Nucl. Part. Phys. Proc.},
 pages = {1883--1890},
 primaryclass = {hep-ex},
 title = {{Search for Heavy Right Handed Neutrinos at the FCC-ee}},
 volume = {273-275},
 year = {2016}
}

@article{Blondel:2021mss,
 archiveprefix = {arXiv},
 author = {Blondel, Alain and de Gouv{\^e}a, Andr{\'e} and Kayser, Boris},
 doi = {10.1103/PhysRevD.104.055027},
 eprint = {2105.06576},
 journal = {Phys. Rev. D},
 number = {5},
 pages = {055027},
 primaryclass = {hep-ph},
 reportnumber = {FERMILAB-PUB-21-227-T},
 title = {{Z-boson decays into Majorana or Dirac heavy neutrinos}},
 volume = {104},
 year = {2021}
}

@article{Blondel:2022qqo,
 archiveprefix = {arXiv},
 author = {Blondel, A. and others},
 doi = {10.3389/fphy.2022.967881},
 eprint = {2203.05502},
 journal = {Front. in Phys.},
 pages = {967881},
 primaryclass = {hep-ex},
 title = {{Searches for long-lived particles at the future FCC-ee}},
 volume = {10},
 year = {2022}
}

@misc{blondel_2025_zv2qx-xk656,
 author = {Blondel, Alain and
Selvaggi, Michele},
 doi = {10.17181/zv2qx-xk656},
 month = {June},
 publisher = {CERN},
 title = {Electroweak factory options for CERN},
 url = {https://doi.org/10.17181/zv2qx-xk656},
 year = {2025}
}

@article{Bodeker:2020ghk,
 archiveprefix = {arXiv},
 author = {Bodeker, Dietrich and Buchmuller, Wilfried},
 doi = {10.1103/RevModPhys.93.035004},
 eprint = {2009.07294},
 journal = {Rev. Mod. Phys.},
 number = {3},
 pages = {035004},
 primaryclass = {hep-ph},
 reportnumber = {DESY 20-141, DESY-20-141},
 title = {{Baryogenesis from the weak scale to the grand unification scale}},
 volume = {93},
 year = {2021}
}

@article{Bondarenko:2019yob,
 archiveprefix = {arXiv},
 author = {Bondarenko, Kyrylo and Boyarsky, Alexey and Ovchynnikov, Maksym and Ruchayskiy, Oleg},
 doi = {10.1007/JHEP08(2019)061},
 eprint = {1902.06240},
 journal = {JHEP},
 pages = {061},
 primaryclass = {hep-ph},
 title = {{Sensitivity of the intensity frontier experiments for neutrino and scalar portals: analytic estimates}},
 volume = {08},
 year = {2019}
}

@article{Boyarsky:2018tvu,
 archiveprefix = {arXiv},
 author = {Boyarsky, A. and Drewes, M. and Lasserre, T. and Mertens, S. and Ruchayskiy, O.},
 doi = {10.1016/j.ppnp.2018.07.004},
 eprint = {1807.07938},
 journal = {Prog. Part. Nucl. Phys.},
 pages = {1--45},
 primaryclass = {hep-ph},
 title = {{Sterile neutrino Dark Matter}},
 volume = {104},
 year = {2019}
}

@article{Boyarsky:2020dzc,
 archiveprefix = {arXiv},
 author = {Boyarsky, Alexey and Ovchynnikov, Maksym and Ruchayskiy, Oleg and Syvolap, Vsevolod},
 doi = {10.1103/PhysRevD.104.023517},
 eprint = {2008.00749},
 journal = {Phys. Rev. D},
 number = {2},
 pages = {023517},
 primaryclass = {hep-ph},
 title = {{Improved big bang nucleosynthesis constraints on heavy neutral leptons}},
 volume = {104},
 year = {2021}
}

@article{Canetti:2012kh,
 archiveprefix = {arXiv},
 author = {Canetti, Laurent and Drewes, Marco and Frossard, Tibor and Shaposhnikov, Mikhail},
 doi = {10.1103/PhysRevD.87.093006},
 eprint = {1208.4607},
 journal = {Phys. Rev. D},
 pages = {093006},
 primaryclass = {hep-ph},
 reportnumber = {TTK-12-05, TUM-HEP-852-12, CAS-KITPC-ITP-368},
 title = {{Dark Matter, Baryogenesis and Neutrino Oscillations from Right Handed Neutrinos}},
 volume = {87},
 year = {2013}
}

@article{Canetti:2012zc,
 archiveprefix = {arXiv},
 author = {Canetti, Laurent and Drewes, Marco and Shaposhnikov, Mikhail},
 doi = {10.1088/1367-2630/14/9/095012},
 eprint = {1204.4186},
 journal = {New J. Phys.},
 pages = {095012},
 primaryclass = {hep-ph},
 reportnumber = {TTK-12-04},
 title = {{Matter and Antimatter in the Universe}},
 volume = {14},
 year = {2012}
}

@article{CEPCStudyGroup:2018ghi,
 archiveprefix = {arXiv},
 author = {Dong, Mingyi and others},
 collaboration = {CEPC Study Group},
 editor = {Guimar{\~a}es da Costa, Jo{\~a}o Barreiro and others},
 eprint = {1811.10545},
 month = {11},
 note = {arXiv: \href{https://arxiv.org/abs/1811.10545}{1811.10545} [hep-ex]},
 primaryclass = {hep-ex},
 reportnumber = {IHEP-CEPC-DR-2018-02, IHEP-EP-2018-01, IHEP-TH-2018-01},
 title = {{CEPC Conceptual Design Report: Volume 2 - Physics {\&} Detector}},
 year = {2018}
}

@article{CEPCStudyGroup:2023quu,
 archiveprefix = {arXiv},
 author = {Abdallah, Waleed and others},
 collaboration = {CEPC Study Group},
 doi = {10.1007/s41605-024-00463-y},
 eprint = {2312.14363},
 journal = {Radiat. Detect. Technol. Methods},
 note = {[Erratum: Radiat.Detect.Technol.Methods 9, 184--192 (2025)]},
 number = {1},
 pages = {1--1105},
 primaryclass = {physics.acc-ph},
 reportnumber = {IHEP-CEPC-DR-2023-01, IHEP-AC-2023-01},
 title = {{CEPC Technical Design Report: Accelerator}},
 volume = {8},
 year = {2024}
}

@article{CEPCStudyGroup:2025kmw,
 archiveprefix = {arXiv},
 author = {Adhya, Souvik Priyam and others},
 collaboration = {CEPC Study Group},
 eprint = {2510.05260},
 month = {10},
 note = {arXiv: \href{https://arxiv.org/abs/2510.05260}{2510.05260} [hep-ex]},
 primaryclass = {hep-ex},
 reportnumber = {IHEP-CEPC-DR-2025-01, IHEP-EP-2025-01},
 title = {{CEPC Technical Design Report - Reference Detector}},
 year = {2025}
}

@article{Chakraborty:2018khw,
 archiveprefix = {arXiv},
 author = {Chakraborty, Sabyasachi and Mitra, Manimala and Shil, Sujay},
 doi = {10.1103/PhysRevD.100.015012},
 eprint = {1810.08970},
 journal = {Phys. Rev. D},
 number = {1},
 pages = {015012},
 primaryclass = {hep-ph},
 title = {{Fat Jet Signature of a Heavy Neutrino at Lepton Collider}},
 volume = {100},
 year = {2019}
}

@article{Chrzaszcz:2020emg,
 archiveprefix = {arXiv},
 author = {Chrz\k{a}szcz, Marcin and Drewes, Marco and Hajer, Jan},
 doi = {10.1140/epjc/s10052-021-09253-y},
 eprint = {2011.01005},
 journal = {Eur. Phys. J. C},
 number = {6},
 pages = {546},
 primaryclass = {hep-ph},
 reportnumber = {CP3-20-48},
 title = {{HECATE: A long-lived particle detector concept for the FCC-ee or CEPC}},
 volume = {81},
 year = {2021}
}

@article{Cowan:2010js,
 archiveprefix = {arXiv},
 author = {Cowan, Glen and Cranmer, Kyle and Gross, Eilam and Vitells, Ofer},
 doi = {10.1140/epjc/s10052-011-1554-0},
 eprint = {1007.1727},
 journal = {Eur. Phys. J. C},
 note = {[Erratum: Eur.Phys.J.C 73, 2501 (2013)]},
 pages = {1554},
 primaryclass = {physics.data-an},
 title = {{Asymptotic formulae for likelihood-based tests of new physics}},
 volume = {71},
 year = {2011}
}

@article{Curtin:2018mvb,
 archiveprefix = {arXiv},
 author = {Curtin, David and others},
 doi = {10.1088/1361-6633/ab28d6},
 eprint = {1806.07396},
 journal = {Rept. Prog. Phys.},
 number = {11},
 pages = {116201},
 primaryclass = {hep-ph},
 reportnumber = {FERMILAB-PUB-18-264-T},
 title = {{Long-Lived Particles at the Energy Frontier: The MATHUSLA Physics Case}},
 volume = {82},
 year = {2019}
}

@article{deBlas:2025gyz,
 archiveprefix = {arXiv},
 author = {de Blas, Jorge and others},
 doi = {10.17181/CERN.35CH.2O2P},
 eprint = {2511.03883},
 month = {11},
 primaryclass = {hep-ex},
 reportnumber = {CERN-ESU-2025-001, CERN-ESU-2025-001},
 title = {{Physics Briefing Book: Input for the 2026 update of the European Strategy for Particle Physics}},
 year = {2025}
}

@article{Deppisch:2015qwa,
 archiveprefix = {arXiv},
 author = {Deppisch, Frank F. and Bhupal Dev, P. S. and Pilaftsis, Apostolos},
 doi = {10.1088/1367-2630/17/7/075019},
 eprint = {1502.06541},
 journal = {New J. Phys.},
 number = {7},
 pages = {075019},
 primaryclass = {hep-ph},
 reportnumber = {MAN-HEP-2014-15},
 title = {{Neutrinos and Collider Physics}},
 volume = {17},
 year = {2015}
}

@article{deVries:2024rfh,
 archiveprefix = {arXiv},
 author = {de Vries, J. and Drewes, M. and Georis, Y. and Klari{\'c}, J. and Plakkot, V.},
 doi = {10.1007/JHEP05(2025)090},
 eprint = {2407.10560},
 journal = {JHEP},
 pages = {090},
 primaryclass = {hep-ph},
 reportnumber = {IRMP-CP3-24-20, ZTF-EP-24-10},
 title = {{Confronting the low-scale seesaw and leptogenesis with neutrinoless double beta decay}},
 volume = {05},
 year = {2025}
}

@article{DiLuzio:2020wdo,
 archiveprefix = {arXiv},
 author = {Di Luzio, Luca and Giannotti, Maurizio and Nardi, Enrico and Visinelli, Luca},
 doi = {10.1016/j.physrep.2020.06.002},
 eprint = {2003.01100},
 journal = {Phys. Rept.},
 pages = {1--117},
 primaryclass = {hep-ph},
 reportnumber = {DESY 20-036, DESY-20-036},
 title = {{The landscape of QCD axion models}},
 volume = {870},
 year = {2020}
}

@article{Dimopoulos:2005ac,
 archiveprefix = {arXiv},
 author = {Dimopoulos, S. and Kachru, S. and McGreevy, J. and Wacker, Jay G.},
 doi = {10.1088/1475-7516/2008/08/003},
 eprint = {hep-th/0507205},
 journal = {JCAP},
 pages = {003},
 reportnumber = {SLAC-PUB-11016, SU-ITP-05-08},
 title = {{N-flation}},
 volume = {08},
 year = {2008}
}

@article{Dine:1982ah,
 author = {Dine, Michael and Fischler, Willy},
 doi = {10.1016/0370-2693(83)90639-1},
 editor = {Srednicki, M. A.},
 journal = {Phys. Lett. B},
 pages = {137--141},
 reportnumber = {UPR-0201T},
 title = {{The Not So Harmless Axion}},
 volume = {120},
 year = {1983}
}

@article{Dodelson:1993je,
 archiveprefix = {arXiv},
 author = {Dodelson, Scott and Widrow, Lawrence M.},
 doi = {10.1103/PhysRevLett.72.17},
 eprint = {hep-ph/9303287},
 journal = {Phys. Rev. Lett.},
 pages = {17--20},
 reportnumber = {FERMILAB-PUB-93-057-A},
 title = {{Sterile-neutrinos as dark matter}},
 volume = {72},
 year = {1994}
}

@article{Drewes:2013gca,
 archiveprefix = {arXiv},
 author = {Drewes, Marco},
 doi = {10.1142/S0218301313300191},
 eprint = {1303.6912},
 journal = {Int. J. Mod. Phys. E},
 pages = {1330019},
 primaryclass = {hep-ph},
 reportnumber = {TUM-HEP-881-13},
 title = {{The Phenomenology of Right Handed Neutrinos}},
 volume = {22},
 year = {2013}
}

@article{Drewes:2016jae,
 archiveprefix = {arXiv},
 author = {Drewes, Marco and Garbrecht, Bjorn and Gueter, Dario and Klaric, Juraj},
 doi = {10.1007/JHEP08(2017)018},
 eprint = {1609.09069},
 journal = {JHEP},
 pages = {018},
 primaryclass = {hep-ph},
 reportnumber = {TUM-HEP-1062-16},
 title = {{Testing the low scale seesaw and leptogenesis}},
 volume = {08},
 year = {2017}
}

@article{Drewes:2016lqo,
 archiveprefix = {arXiv},
 author = {Drewes, Marco and Eijima, Shintaro},
 doi = {10.1016/j.physletb.2016.09.054},
 eprint = {1606.06221},
 journal = {Phys. Lett. B},
 pages = {72--79},
 primaryclass = {hep-ph},
 reportnumber = {TUM-HEP-1049-16},
 title = {{Neutrinoless double $\beta$ decay and low scale leptogenesis}},
 volume = {763},
 year = {2016}
}

@article{Drewes:2019fou,
 archiveprefix = {arXiv},
 author = {Drewes, Marco and Hajer, Jan},
 doi = {10.1007/JHEP02(2020)070},
 eprint = {1903.06100},
 journal = {JHEP},
 pages = {070},
 primaryclass = {hep-ph},
 reportnumber = {CP3-19-11},
 title = {{Heavy Neutrinos in displaced vertex searches at the LHC and HL-LHC}},
 volume = {02},
 year = {2020}
}

@article{Drewes:2019mhg,
 archiveprefix = {arXiv},
 author = {Drewes, Marco},
 eprint = {1904.11959},
 month = {4},
 note = {arXiv: \href{https://arxiv.org/abs/1904.11959}{1904.11959} [hep-ph]},
 primaryclass = {hep-ph},
 reportnumber = {CP3-19-20},
 title = {{On the Minimal Mixing of Heavy Neutrinos}},
 year = {2019}
}

@article{Drewes:2019vjy,
 archiveprefix = {arXiv},
 author = {Drewes, Marco and Giammanco, Andrea and Hajer, Jan and Lucente, Michele},
 doi = {10.1103/PhysRevD.101.055002},
 eprint = {1905.09828},
 journal = {Phys. Rev. D},
 number = {5},
 pages = {055002},
 primaryclass = {hep-ph},
 reportnumber = {CP3-19-26},
 title = {{New long-lived particle searches in heavy-ion collisions at the LHC}},
 volume = {101},
 year = {2020}
}

@article{Drewes:2021nqr,
 archiveprefix = {arXiv},
 author = {Drewes, Marco and Georis, Yannis and Klari\'c, Juraj},
 doi = {10.1103/PhysRevLett.128.051801},
 eprint = {2106.16226},
 journal = {Phys. Rev. Lett.},
 number = {5},
 pages = {051801},
 primaryclass = {hep-ph},
 reportnumber = {CP3-21-43},
 title = {{Mapping the Viable Parameter Space for Testable Leptogenesis}},
 volume = {128},
 year = {2022}
}

@article{Drewes:2022akb,
 archiveprefix = {arXiv},
 author = {Drewes, Marco and Klari{\'c}, Juraj and L{\'o}pez-Pav{\'o}n, Jacobo},
 doi = {10.1140/epjc/s10052-022-11100-7},
 eprint = {2207.02742},
 journal = {Eur. Phys. J. C},
 number = {12},
 pages = {1176},
 primaryclass = {hep-ph},
 title = {{New benchmark models for heavy neutral lepton searches}},
 volume = {82},
 year = {2022}
}

@article{Drewes:2022rsk,
 archiveprefix = {arXiv},
 author = {Drewes, Marco},
 doi = {10.22323/1.414.0608},
 eprint = {2210.17110},
 journal = {PoS},
 pages = {608},
 primaryclass = {hep-ph},
 reportnumber = {IRMP-CP3-22-52},
 title = {{Distinguishing Dirac and Majorana Heavy Neutrinos at Lepton Colliders}},
 volume = {ICHEP2022},
 year = {2022}
}

@article{Drewes:2024bla,
 archiveprefix = {arXiv},
 author = {Drewes, Marco and Georis, Yannis and Klari{\'c}, Juraj and Wendels, Antony},
 doi = {10.1007/JHEP03(2025)176},
 eprint = {2407.13620},
 journal = {JHEP},
 pages = {176},
 primaryclass = {hep-ph},
 reportnumber = {IRMP-CP3-24-21, ZTF-EP-24-11},
 title = {{On the collider-testability of the type-I seesaw model with 3 right-handed neutrinos}},
 volume = {03},
 year = {2025}
}

@article{Drewes:2024pad,
 archiveprefix = {arXiv},
 author = {Drewes, Marco and Georis, Yannis and Hagedorn, Claudia and Klaric, Juraj},
 eprint = {2412.10254},
 month = {12},
 note = {arXiv: \href{https://arxiv.org/abs/2412.10254}{2412.10254} [hep-ph]},
 primaryclass = {hep-ph},
 title = {{Low-scale seesaw with flavour and CP symmetries {\textendash} from colliders to leptogenesis}},
 year = {2024}
}

@inproceedings{Drewes:2025ohy,
 archiveprefix = {arXiv},
 author = {Drewes, Marco and Shaposhnikova, Elena and Shaposhnikov, Mikhail},
 eprint = {2503.17081},
 month = {3},
 note = {arXiv: \href{https://arxiv.org/abs/2503.17081}{2503.17081} [hep-ex]},
 primaryclass = {hep-ex},
 title = {{A Possible Future Use of the LHC Tunnel}},
 year = {2025}
}

@article{DUNE:2015lol,
 archiveprefix = {arXiv},
 author = {Acciarri, R. and others},
 collaboration = {DUNE},
 eprint = {1512.06148},
 month = {12},
 note = {arXiv: \href{https://arxiv.org/abs/1512.06148}{1512.06148} [physics.ins-det]},
 primaryclass = {physics.ins-det},
 reportnumber = {FERMILAB-DESIGN-2016-02},
 title = {{Long-Baseline Neutrino Facility (LBNF) and Deep Underground Neutrino Experiment (DUNE)}: {Conceptual Design Report, Volume 2: The Physics Program for DUNE at LBNF}},
 year = {2015}
}

@article{DUNE:2020ypp,
 archiveprefix = {arXiv},
 author = {Abi, Babak and others},
 collaboration = {DUNE},
 eprint = {2002.03005},
 month = {2},
 note = {arXiv: \href{https://arxiv.org/abs/2002.03005}{2002.03005} [hep-ex]},
 primaryclass = {hep-ex},
 reportnumber = {FERMILAB-PUB-20-025-ND, FERMILAB-DESIGN-2020-02},
 title = {{Deep Underground Neutrino Experiment (DUNE), Far Detector Technical Design Report, Volume II: DUNE Physics}},
 year = {2020}
}

@article{Esteban:2024eli,
 archiveprefix = {arXiv},
 author = {Esteban, Ivan and Gonzalez-Garcia, M. C. and Maltoni, Michele and Martinez-Soler, Ivan and Pinheiro, Jo{\~a}o Paulo and Schwetz, Thomas},
 doi = {10.1007/JHEP12(2024)216},
 eprint = {2410.05380},
 journal = {JHEP},
 pages = {216},
 primaryclass = {hep-ph},
 reportnumber = {IFT-UAM/CSIC-24-140, YITP-SB-2024-24, IPPP/24/64, IPPP/24/64, IFT-UAM/CSIC-24-140, YITP-SB-2024-24},
 title = {{NuFit-6.0: updated global analysis of three-flavor neutrino oscillations}},
 volume = {12},
 year = {2024}
}

@article{FCC:2018evy,
 author = {Abada, A. and others},
 collaboration = {FCC},
 doi = {10.1140/epjst/e2019-900045-4},
 journal = {Eur. Phys. J. ST},
 number = {2},
 pages = {261--623},
 reportnumber = {CERN-ACC-2018-0057},
 title = {{FCC-ee: The Lepton Collider}: {Future Circular Collider Conceptual Design Report Volume 2}},
 volume = {228},
 year = {2019}
}

@article{FCC:2025lpp,
 archiveprefix = {arXiv},
 author = {Benedikt, M. and others},
 collaboration = {FCC},
 doi = {10.17181/CERN.9DKX.TDH9},
 eprint = {2505.00272},
 month = {4},
 primaryclass = {hep-ex},
 reportnumber = {CERN-FCC-PHYS-2025-0002},
 title = {{Future Circular Collider Feasibility Study Report: Volume 1, Physics, Experiments, Detectors}},
 year = {2025}
}

@article{Freese:1990rb,
 author = {Freese, Katherine and Frieman, Joshua A. and Olinto, Angela V.},
 doi = {10.1103/PhysRevLett.65.3233},
 journal = {Phys. Rev. Lett.},
 pages = {3233--3236},
 reportnumber = {FERMILAB-PUB-90-177-A},
 title = {{Natural inflation with pseudo - Nambu-Goldstone bosons}},
 volume = {65},
 year = {1990}
}

@article{Fukugita:1986hr,
 author = {Fukugita, M. and Yanagida, T.},
 doi = {10.1016/0370-2693(86)91126-3},
 journal = {Phys. Lett. B},
 pages = {45--47},
 reportnumber = {RIFP-641},
 title = {{Baryogenesis Without Grand Unification}},
 volume = {174},
 year = {1986}
}

@article{Gell-Mann:1979vob,
 archiveprefix = {arXiv},
 author = {Gell-Mann, Murray and Ramond, Pierre and Slansky, Richard},
 eprint = {1306.4669},
 journal = {Conf. Proc. C},
 pages = {315--321},
 primaryclass = {hep-th},
 reportnumber = {PRINT-80-0576},
 title = {{Complex Spinors and Unified Theories}},
 volume = {790927},
 year = {1979}
}

@article{Ghiglieri:2020ulj,
 archiveprefix = {arXiv},
 author = {Ghiglieri, J. and Laine, M.},
 doi = {10.1088/1475-7516/2020/07/012},
 eprint = {2004.10766},
 journal = {JCAP},
 pages = {012},
 primaryclass = {hep-ph},
 title = {{Sterile neutrino dark matter via coinciding resonances}},
 volume = {07},
 year = {2020}
}

@article{Glashow:1979nm,
 author = {Glashow, S. L.},
 doi = {10.1007/978-1-4684-7197-7_15},
 editor = {L{\'e}vy, Maurice and Basdevant, Jean-Louis and Speiser, David and Weyers, Jacques and Gastmans, Raymond and Jacob, Maurice},
 journal = {NATO Sci. Ser. B},
 pages = {687},
 reportnumber = {HUTP-79-A059},
 title = {{The Future of Elementary Particle Physics}},
 volume = {61},
 year = {1980}
}

@article{Hernandez:2016kel,
 archiveprefix = {arXiv},
 author = {Hern{\'a}ndez, P. and Kekic, M. and L{\'o}pez-Pav{\'o}n, J. and Racker, J. and Salvado, J.},
 doi = {10.1007/JHEP08(2016)157},
 eprint = {1606.06719},
 journal = {JHEP},
 pages = {157},
 primaryclass = {hep-ph},
 title = {{Testable Baryogenesis in Seesaw Models}},
 volume = {08},
 year = {2016}
}

@article{Hernandez:2022ivz,
 archiveprefix = {arXiv},
 author = {Hernandez, Pilar and Lopez-Pavon, Jacobo and Rius, Nuria and Sandner, Stefan},
 doi = {10.1007/JHEP12(2022)012},
 eprint = {2207.01651},
 journal = {JHEP},
 pages = {012},
 primaryclass = {hep-ph},
 reportnumber = {IFIC/22-20, FTUV-22-0704.1758},
 title = {{Bounds on right-handed neutrino parameters from observable leptogenesis}},
 volume = {12},
 year = {2022}
}

@article{IDEAStudyGroup:2025gbt,
 archiveprefix = {arXiv},
 author = {Abbrescia, M. and others},
 collaboration = {IDEA Study Group},
 eprint = {2502.21223},
 month = {2},
 note = {arXiv: \href{https://arxiv.org/abs/2502.21223}{2502.21223} [physics.ins-det]},
 primaryclass = {physics.ins-det},
 reportnumber = {FERMILAB-PUB-25-0189-PPD},
 title = {{The IDEA detector concept for FCC-ee}},
 year = {2025}
}

@article{Kim:2004rp,
 archiveprefix = {arXiv},
 author = {Kim, Jihn E. and Nilles, Hans Peter and Peloso, Marco},
 doi = {10.1088/1475-7516/2005/01/005},
 eprint = {hep-ph/0409138},
 journal = {JCAP},
 pages = {005},
 title = {{Completing natural inflation}},
 volume = {01},
 year = {2005}
}

@article{LoChiatto:2024guj,
 archiveprefix = {arXiv},
 author = {Lo Chiatto, Prisco and Yu, Felix},
 doi = {10.1103/PhysRevD.111.035001},
 eprint = {2405.03396},
 journal = {Phys. Rev. D},
 number = {3},
 pages = {035001},
 primaryclass = {hep-ph},
 reportnumber = {MITP-23-057},
 title = {{Consistent electroweak phenomenology of a nearly degenerate Z' boson}},
 volume = {111},
 year = {2025}
}

@article{Mimasu:2014nea,
 archiveprefix = {arXiv},
 author = {Mimasu, Ken and Sanz, Ver{\'o}nica},
 doi = {10.1007/JHEP06(2015)173},
 eprint = {1409.4792},
 journal = {JHEP},
 pages = {173},
 primaryclass = {hep-ph},
 title = {{ALPs at Colliders}},
 volume = {06},
 year = {2015}
}

@article{Minkowski:1977sc,
 author = {Minkowski, Peter},
 doi = {10.1016/0370-2693(77)90435-X},
 journal = {Phys. Lett. B},
 pages = {421--428},
 reportnumber = {Print-77-0182 (BERN)},
 title = {{$\mu \to e\gamma$ at a Rate of One Out of $10^{9}$ Muon Decays?}},
 volume = {67},
 year = {1977}
}

@article{Mohapatra:1979ia,
 author = {Mohapatra, Rabindra N. and Senjanovic, Goran},
 doi = {10.1103/PhysRevLett.44.912},
 journal = {Phys. Rev. Lett.},
 pages = {912},
 reportnumber = {MDDP-TR-80-060, MDDP-PP-80-105, CCNY-HEP-79-10},
 title = {{Neutrino Mass and Spontaneous Parity Nonconservation}},
 volume = {44},
 year = {1980}
}

@article{Peccei:1977hh,
 author = {Peccei, R. D. and Quinn, Helen R.},
 doi = {10.1103/PhysRevLett.38.1440},
 journal = {Phys. Rev. Lett.},
 pages = {1440--1443},
 reportnumber = {ITP-568-STANFORD},
 title = {{CP Conservation in the Presence of Instantons}},
 volume = {38},
 year = {1977}
}

@article{Peccei:1977ur,
 author = {Peccei, R. D. and Quinn, Helen R.},
 doi = {10.1103/PhysRevD.16.1791},
 journal = {Phys. Rev. D},
 pages = {1791--1797},
 reportnumber = {ITP-572-STANFORD},
 title = {{Constraints Imposed by CP Conservation in the Presence of Instantons}},
 volume = {16},
 year = {1977}
}

@article{Planck:2018vyg,
 archiveprefix = {arXiv},
 author = {Aghanim, N. and others},
 collaboration = {Planck},
 doi = {10.1051/0004-6361/201833910},
 eprint = {1807.06209},
 journal = {Astron. Astrophys.},
 note = {[Erratum: Astron.Astrophys. 652, C4 (2021)]},
 pages = {A6},
 primaryclass = {astro-ph.CO},
 title = {{Planck 2018 results. VI. Cosmological parameters}},
 volume = {641},
 year = {2020}
}

@article{Polesello:2025gwj,
 archiveprefix = {arXiv},
 author = {Polesello, Giacomo},
 doi = {10.1007/JHEP06(2025)239},
 eprint = {2502.08411},
 journal = {JHEP},
 pages = {239},
 primaryclass = {hep-ph},
 title = {{Sensitivity of the FCC-ee to decay of an axion-like particle into two photons}},
 volume = {06},
 year = {2025}
}

@article{Preskill:1982cy,
 author = {Preskill, John and Wise, Mark B. and Wilczek, Frank},
 doi = {10.1016/0370-2693(83)90637-8},
 editor = {Srednicki, M. A.},
 journal = {Phys. Lett. B},
 pages = {127--132},
 reportnumber = {HUTP-82-A048, NSF-ITP-82-103},
 title = {{Cosmology of the Invisible Axion}},
 volume = {120},
 year = {1983}
}

@article{RebelloTeles:2023uig,
 archiveprefix = {arXiv},
 author = {Rebello Teles, Patricia and d'Enterria, David and Gon{\c{c}}alves, Victor P. and Martins, Daniel E.},
 doi = {10.1103/PhysRevD.109.055003},
 eprint = {2310.17270},
 journal = {Phys. Rev. D},
 number = {5},
 pages = {055003},
 primaryclass = {hep-ex},
 title = {{Searches for axionlike particles via {\ensuremath{\gamma}}{\ensuremath{\gamma}} fusion at future e+e- colliders}},
 volume = {109},
 year = {2024}
}

@article{Schechter:1980gr,
 author = {Schechter, J. and Valle, J. W. F.},
 doi = {10.1103/PhysRevD.22.2227},
 journal = {Phys. Rev. D},
 pages = {2227},
 reportnumber = {SU-4217-167, COO-3533-167},
 title = {{Neutrino Masses in SU(2) x U(1) Theories}},
 volume = {22},
 year = {1980}
}

@article{Shen:2022ffi,
 archiveprefix = {arXiv},
 author = {Shen, Yin-Fa and Ding, Jian-Nan and Qin, Qin},
 doi = {10.1140/epjc/s10052-022-10301-4},
 eprint = {2201.05831},
 journal = {Eur. Phys. J. C},
 number = {5},
 pages = {398},
 primaryclass = {hep-ph},
 title = {{Monojet search for heavy neutrinos at future Z-factories}},
 volume = {82},
 year = {2022}
}

@article{SHiP:2015vad,
 archiveprefix = {arXiv},
 author = {Anelli, M. and others},
 collaboration = {SHiP},
 eprint = {1504.04956},
 month = {4},
 note = {arXiv: \href{https://arxiv.org/abs/1504.04956}{1504.04956} [physics.ins-det]},
 primaryclass = {physics.ins-det},
 reportnumber = {CERN-SPSC-2015-016, SPSC-P-350},
 title = {{A facility to Search for Hidden Particles (SHiP) at the CERN SPS}},
 year = {2015}
}

@article{SHiP:2025ows,
 archiveprefix = {arXiv},
 author = {Albanese, R. and others},
 collaboration = {SHiP, HI-ECN3 Project Team},
 eprint = {2504.06692},
 month = {4},
 note = {arXiv: \href{https://arxiv.org/abs/2504.06692}{2504.06692} [hep-ex]},
 primaryclass = {hep-ex},
 title = {{SHiP experiment at the SPS Beam Dump Facility}},
 year = {2025}
}

@article{Silverstein:2008sg,
 archiveprefix = {arXiv},
 author = {Silverstein, Eva and Westphal, Alexander},
 doi = {10.1103/PhysRevD.78.106003},
 eprint = {0803.3085},
 journal = {Phys. Rev. D},
 pages = {106003},
 primaryclass = {hep-th},
 reportnumber = {SU-ITP-08-07, SLAC-PUB-13183},
 title = {{Monodromy in the CMB: Gravity Waves and String Inflation}},
 volume = {78},
 year = {2008}
}

@article{Tian:2022rsi,
 archiveprefix = {arXiv},
 author = {Tian, Minglun and Wang, Zeren Simon and Wang, Kechen},
 eprint = {2201.08960},
 month = {1},
 note = {arXiv: \href{https://arxiv.org/abs/2201.08960}{2201.08960} [hep-ph]},
 primaryclass = {hep-ph},
 title = {{Search for long-lived axions with far detectors at future lepton colliders}},
 year = {2022}
}

@article{Wang:2019xvx,
 archiveprefix = {arXiv},
 author = {Wang, Zeren Simon and Wang, Kechen},
 doi = {10.1103/PhysRevD.101.075046},
 eprint = {1911.06576},
 journal = {Phys. Rev. D},
 number = {7},
 pages = {075046},
 primaryclass = {hep-ph},
 reportnumber = {APCTP Pre2019-024},
 title = {{Physics with far detectors at future lepton colliders}},
 volume = {101},
 year = {2020}
}

@article{Weinberg:1977ma,
 author = {Weinberg, Steven},
 doi = {10.1103/PhysRevLett.40.223},
 journal = {Phys. Rev. Lett.},
 pages = {223--226},
 reportnumber = {HUTP-77/A074},
 title = {{A New Light Boson?}},
 volume = {40},
 year = {1978}
}

@article{Wilczek:1977pj,
 author = {Wilczek, Frank},
 doi = {10.1103/PhysRevLett.40.279},
 journal = {Phys. Rev. Lett.},
 pages = {279--282},
 reportnumber = {Print-77-0939 (COLUMBIA)},
 title = {{Problem of Strong  $P$  and  $T$  Invariance in the Presence of Instantons}},
 volume = {40},
 year = {1978}
}

@misc{wilkinson_2025_nf05j-xsq05,
 author = {Wilkinson, Guy and
Monteil, Stephane and
F. Kamenik, Jernej and
Lusiani, Alberto},
 doi = {10.17181/nf05j-xsq05},
 month = {March},
 publisher = {CERN},
 title = {Prospects in flavour physics at the FCC},
 url = {https://doi.org/10.17181/nf05j-xsq05},
 year = {2025}
}

@article{Yanagida:1980xy,
 author = {Yanagida, Tsutomu},
 doi = {10.1143/PTP.64.1103},
 journal = {Prog. Theor. Phys.},
 pages = {1103},
 reportnumber = {TU-80-208},
 title = {{Horizontal Symmetry and Masses of Neutrinos}},
 volume = {64},
 year = {1980}
}
\end{document}